\documentclass[10pt,journal,compsoc]{IEEEtran}
\makeatletter
\long\def\@makecaption#1#2{\ifx\@captype\@IEEEtablestring%
\footnotesize\begin{center}{\normalfont\footnotesize #1}\\
{\normalfont\footnotesize\scshape #2}\end{center}%
\@IEEEtablecaptionsepspace
\else
\@IEEEfigurecaptionsepspace
\setbox\@tempboxa\hbox{\normalfont\footnotesize {#1.}~~ #2}%
\ifdim \wd\@tempboxa >\hsize%
\setbox\@tempboxa\hbox{\normalfont\footnotesize {#1.}~~ }%
\parbox[t]{\hsize}{\normalfont\footnotesize \noindent\unhbox\@tempboxa#2}%
\else
\hbox to\hsize{\normalfont\footnotesize\hfil\box\@tempboxa\hfil}\fi\fi}
\makeatother
\ifCLASSOPTIONcompsoc
  \usepackage[nocompress]{cite}
\else
  \usepackage{cite}
\fi
\ifCLASSINFOpdf
  \usepackage[pdftex]{graphicx}
  \graphicspath{{images/}{Authors/}}
\else
\fi
\usepackage{amsmath}
\usepackage{algorithmic}

\usepackage{array}

\ifCLASSOPTIONcompsoc
  \usepackage[caption=false,font=footnotesize,labelfont=sf,textfont=sf]{subfig}
\else
  \usepackage[caption=false,font=footnotesize]{subfig}
\fi
\begin{document}
%
\title{Bayesian Overlapping Community Detection in Dynamic Networks}
%
%
%
%

\author{Mahsa~Ghorbani,
        Hamid~R.~Rabiee,
        and~Ali~Khodadadi,
        \IEEEcompsocitemizethanks{\IEEEcompsocthanksitem H. R. Rabiee, M. Ghorbani and A. Khodadadi are with the Department
of Computer Engineering, Sharif University of Technology, Tehran, Iran.\protect\\
E-mail: rabiee@sharif.edu, \{mghorbani,khodadadi\}@ce.sharif.edu
}
}

%
%

\markboth{Journal of \LaTeX\ Class Files,~Vol.~3, No.~2, May~2016}%
{Shell \MakeLowercase{\textit{et al.}}: Bare Demo of IEEEtran.cls for Computer Society Journals}
%



\IEEEtitleabstractindextext{%
\begin{abstract}
Detecting community structures in social networks has gained considerable attention in recent years. However, lack of prior knowledge about the number of communities, and their overlapping nature have made community detection a challenging problem. Moreover, many of the existing methods only consider static networks, while most of real world networks are dynamic and evolve over time. Hence, finding consistent overlapping communities in dynamic networks without any prior knowledge about the number of communities is still an interesting open research problem. In this paper, we present an overlapping community detection method for dynamic networks called Dynamic Bayesian Overlapping Community Detector (DBOCD). DBOCD assumes that in every snapshot of network, overlapping parts of communities are dense areas and utilizes link communities instead of common node communities. Using Recurrent Chinese Restaurant Process and community structure of the network in the last snapshot, DBOCD simultaneously extracts the number of communities and soft community memberships of nodes while maintaining the consistency of communities over time. We evaluated DBOCD on both synthetic and real dynamic datasets to assess its ability to find overlapping communities in different types of network evolution. The results show that DBOCD outperforms the recent state of the art dynamic community detection methods.
\end{abstract}

}

\maketitle

\IEEEdisplaynontitleabstractindextext

%
\IEEEpeerreviewmaketitle

\IEEEraisesectionheading{\section{Introduction}\label{sec:introduction}}
\IEEEPARstart{O}{ne} of the most important research problems in network science is the identification of the community structure of networks. In a network, a community is a group of nodes with many intra-group connections and a few ones toward outside \cite{girvan2002community}. Identifying communities has many real world applications. Suggesting items in recommender systems, detecting spy and terrorist groups, and predicting future links between members of a social network are some examples where community detection algorithms have been utilized \cite{sahebi2011community,li2014clustering}.
\par The community detection problem has been investigated comprehensively in the last decade using different approaches and assumptions. Traditional methods assume the number of communities is known a priori, and every node belongs to exactly one community. However, in real world a node can belong to many groups, simultaneously \cite{fortunato2010community}, and often, we do not have any information about the true number of communities. Furthermore, in many community detection studies, the network is considered to be static, which means the network is fixed and does not change over time. However, in real world the networks and communities change constantly over time, because members and connections are added and removed from the network \cite{palla2007quantifying}.
Recently, community detection in dynamic networks has gained some attention \cite{xu2014adaptive,yang2011detecting,lin2008facetnet}. Preserving the consistency of communities over time is the main challenge of these methods and due to scalability issues and large number of parameters, many of them concentrate on disjoint communities. Hence, introducing a method to find overlapping communities in dynamic networks without any knowledge about the number of communities with appropriate complexity is an open research problem.

\par In this paper, we study the overlapping community detection  in dynamic networks without any prior knowledge about the true number of communities. 
To make the model simple to understand, we divide the proposed generative model into two steps. In the first step, we describe a generative model for a static network based on link communities and in the second step, we extend the generative model for dynamic networks. Afterwards, to infer the model parameters and extract the community memberships of nodes, we apply statistical inference methods on observed snapshots of an evolving network. 
The proposed method is called Dynamic Bayesian Overlapping Community Detector (DBOCD). DBOCD is able to discover consistent overlapping communities and their numbers simultaneously, in polynomial time by using the Recurrent Chinese Restaurant Process (RCRP) as a prior knowledge, and the adjacency matrix as observations in every time snapshots of network. The main contributions of the proposed work are: 
\begin{itemize}
\item Estimating the number of communities automatically in each time snapshot while preserving robustness and attention to rich-get-richer phenomena in networks.
\item Decreasing time complexity by using link partitioning instead of the common node partitioning.
\item Handling different types of community and network evolution.
\item Estimating the soft memberships of nodes in communities (detecting overlapping communities).
\item Providing a theoretical analysis of time complexity of the proposed method.
\end{itemize}
Experimental results on synthetic and real datasets indicate the ability of the proposed method in capturing different types of evolution compared to the recent methods. 
\par The rest of this paper is organized as follows. The related works are reviewed in Section \ref{related:sec}. Section \ref{background:sec} provides the preliminary concepts that are being used in the proposed method. The proposed static method and its dynamic extension are explained in Section \ref{method:sec}. The experimental results on synthetic and real datasets are provided in Section \ref{experiment:sec}. Finally, conclusions and future works are discussed in Section \ref{conclusion:sec}.
\section{Related Works}
\label{related:sec}
Communities and their identification in static networks have been studied extensively in recent years.
According to \cite{xie2013overlapping}, the previous studies can be divided into 4 categories. (1) \textbf{Optimization methods}, which are based on maximizing the definition of a good community structure in graphs \cite{reichardt2006statistical,blondel2008fast}, (2) \textbf{Seed expansion methods}, that select some nodes as core of communities and expand them to cover the graph \cite{baumes2005finding,lancichinetti2011finding}. (3) \textbf{Clique based methods}, that define communities as complete or near complete subgraphs \cite{adamcsek2006cfinder,kumpula2008sequential}. (4) \textbf{Probabilistic approaches}, which propose a generative model for network generation and fit parameters to find the best structure of communities  \cite{airoldi2009mixed,psorakis2011overlapping}. 
\par In another view, community detection methods can be divided into node based and edge based methods. In node based methods, the community membership is assigned explicitly to nodes. However, in edge based methods, the links between nodes belong to communities and the membership of nodes are implicitly extracted from information about the link communities\cite{he2014link,ahn2010link,ren2009simple}.
\par Traditional studies concentrate on static networks, while recent methods try to find and track the evolution of communities in dynamic networks. These studies usually consider the dynamic network as a set of network snapshots in discrete times and try to find and track the communities using these snapshots of network. The main concern of these works is to find consistent communities over time using some constraints on changing the community membership of nodes in adjacent snapshots.
The authors in \cite{lin2008facetnet} proposed FacetNet, which models community memberships as a latent dimension in networks. FacetNet uses Kullback-Leibler divergence to limit the changes of communities in two adjacent snapshots. FacetNet needs the number of communities as input at each time step. The authors in \cite{yang2011detecting} proposed Dynamic Stochastic Block Model (DSBM), which is a dynamic extension of SBM \cite{leinhardt1976local}.
DSBM detects non-overlapping communities in networks with constant number of nodes and communities over time and also needs the number of communities. The authors in \cite{xu2014adaptive} proposed the AFFECT algorithm to discover a noise-free network utilizing observed network and its history. After discovering a noise-free network, static non-overlapping community detection methods are used at each snapshot. The authors in \cite{tantipathananandh2011finding} proposed the SDP method which considers penalties for node membership alteration to control difference between discovered communities at two snapshots. SDP is incapable of discovering the number of communities and has too many parameters. 
DPLA+\cite{liu2015label} is a label-propagation based method for detecting overlapping and non-overlapping communities in dynamic networks, which the community membership of each node is based on the membership of its neighbors and their relative importance. A weighted mean between the importance of a node in the previous snapshot and the current snapshot is calculated to find consistent communities over time. Label-propagation based methods are usually sensitive to density of the network and the number of iterations.
\par Most of the community detection methods in static or dynamic networks need the number of communities to be known a priori. Using non-parametric methods is a new approach for estimating the number of communities in static \cite{morup2012bayesian,whang2013stochastic} or dynamic networks \cite{tang2011dynamic}.
Due to scalability issues and large number of parameters, there is no prior work that is able to find the number of communities and the soft memberships (overlapping communities), simultaneously in dynamic networks. In this paper, we have concentrated on the overlapping community detection in dynamic networks without any prior knowledge about the number of communities with polynomial complexity. The proposed method is able to find the soft community memberships, the number of communities, and also preserve the consistency of communities over time.
 
\section{Background}
In this paper, we use non-parametric models to find the number of communities. In this section, we describe two non-parametric models which have been  used in our model.
\label{background:sec}
\\\textbf{Chinese Restaurant Process (CRP):}
Chinese restaurant process is a non-parametric process for partitioning data into an unknown number of non-overlapping groups. Consider a Chinese restaurant with infinite number of tables. The first customer enters and selects a table at random. The $i$-th customer selects a table that has been chosen previously with probability $\frac{n_k}{i-1+\alpha}$ and a new table with probability $\frac{\alpha}{i-1+\alpha}$, where $n_k$ is the number of customers around table $k$ and $\alpha$ is the hyper-parameter of model controlling the rate at which a new table is added. In CRP, every table is a group and every customer is an observation of data \cite{pitman2002combinatorial}. We can see that a table with plenty of customers has more chance to be chosen by later customers which is compatible with rich-get-richer phenomena in social networks.
\\\textbf{Recurrent Chinese Restaurant Process (RCRP):}
There are many non-parametric approaches to model a system over time \cite{blei2011distance,ahmed2008dynamic}, one of the most popular methods is RCRP \cite{ahmed2008dynamic}. RCRP is the timed extension of the CRP model. In RCRP, each snapshot is a day. At the end of the day, restaurant becomes empty and the popularity of tables are analyzed. The main idea is that a popular table on a day remains popular for the next day. At day $t$ ($t\neq 1$), the $i$-th customer can select table $k$ which is used in the previous day ($t-1$) with a probability proportional to $\frac{n_{k,t-1}+n_{k,t}^{(i)}}{N_{t-1}+i-1+\alpha}$, where $n_{k,t-1}$ is the number of customers using table $k$ at day $t-1$, $N_{t-1}$ is the total number of customers at day $t-1$, and $n_{k,t}^{(i)}$ is the number of customers using table $k$ at day $t$ at the time of the arrival of customer $i$. If table $k$ has not been chosen before, $n_{k,t}^{(i)}$ will be set to zero. Customer $i$, can choose a new table that is not used in the previous day. The
probability of selecting a new table is $\frac{\alpha}{N_{t-1}+i-1+\alpha}$ \cite{ahmed2008dynamic}.
For a thorough study of CRP and RCRP, we refer the readers to \cite{pitman2002combinatorial, ahmed2008dynamic}.
In the next section, we utilize CRP to propose a static model based on link communities for overlapping community detection. Then RCRP will be used to extend the static model to dynamic networks.
\section{Generative Model}
\label{method:sec}
As we mentioned before, our goal is to introduce a method for detecting overlapping communities in a network that evolves over time, considering the consistency of communities in the network snapshots. 
We first describe the static version of our generative model, then we extend it to dynamic networks. It is worth noting that these steps are taken for making the model simple to understand, and our aim is to determine the overlapping communities in dynamic networks.
\subsection{Generative Model for Static Networks}
\begin{figure}[!t]
\centering
\includegraphics[width=0.25\textwidth]{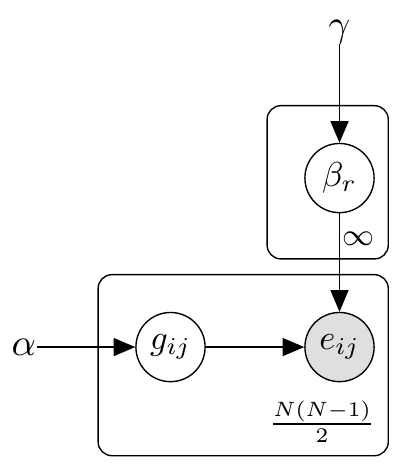}
\caption{Graphical model for static network generation in BOCD from latent parameters $B$ and $G$ with hyper-parameters $\alpha$ and $\gamma$.}
\label{graphicalModelFig}
\end{figure}

In overlapping community detection problems with unknown number of communities, while the number of communities may vary, nodes can belong to multiple communities with various degrees of memberships. These assumptions make the space of possible solutions very large. Previous methods, like IBP \cite{griffiths2011indian}, have used Bayesian non-parametric methods to simultaneously find the number of communities, and soft memberships of nodes in communities \cite{whang2013stochastic}. However, because of complexity of the solution space, such methods usually have exponential time complexity. To solve this problem, we use link communities and assume that in addition to the nodes of a network, links can belong to communities, and link communities are non-overlapping. By this assumption, we can find the number of link communities and their members by simple non-overlapping Bayesian non-parametric methods. Then, by using the connected links to a node, the degree of membership for every node can be determined. Therefore, by utilizing this approach, while the space of possible solutions is more confined than before, we can find the number of communities and the soft memberships of nodes, simultaneously. 
Our basic idea for static networks is to use non-overlapping communities of links, and CRP as a prior on link communities to find the number of communities and the soft membership of nodes in communities simultaneously. \\
The link based communities have been used previously in literature \cite{ren2009simple}, and we use the same notation to describe the model. Assume a network with $N$ nodes and an unknown number of communities. We define $\beta_r$ for each community, which is a vector of size $N$ and must satisfy the constraint that $\sum_{i=1}^N \beta_{ir} = 1$. $\beta_{ir}$ denotes the importance of node $i$ in community $r$. $B$ denotes the set of all $\beta_r$s.  We assume that each edge belongs to one community and as much as two nodes are more important in community $r$, the edge between them has higher probability to belong to the community $r$. $e_{ij}$ is a binary variable indicating the edge between nodes $i$ and $j$, and the variable $g_{ij}$ denotes the community membership of edge $e_{ij}$. We show the adjacency matrix of graph with $A$, and the set of all $g_{ij}$s with $G$.
Prior distributions on variables of model are assigned as:
\begin{equation}
G\sim CRP\left( \alpha  \right)  
\label{GPriorDist}
\end{equation}
\begin{equation}
\beta_{r} \sim Dir\left( {\gamma } \right)  
\label{betaPriorDist}
\end{equation}
Knowing the group membership of edge and $B$, the probability of link generation is:
\begin{equation}
{e}_{ij}|(g_{ij}=r, B) \sim Bernoulli~ ( \beta_{ir}\beta_{jr})
\label{EPriorDist}
\end{equation}
The graphical model for static network generation is depicted in Fig.\ref{graphicalModelFig}. 
Based  on the proposed generative model, the joint distribution over variables factorizes as Eq. \ref{factorizedJointDistEq},
\begin{equation}
\begin{split}
p(G,B ,A|\alpha ,\gamma )=p(A|B ,G) p( G| \alpha) p(B|\gamma)= \\
\prod_{i=1}^{N} \prod_{j=1}^{N} p({e}_{ij}|{g}_{ij},\beta_{g_{ij}}) p(G|\alpha) \prod_{i=1}^K p(\beta_{i}|\gamma)
\end{split}
\label{factorizedJointDistEq}
\end{equation}
where, K is the number of communities. Substituting \ref{GPriorDist}, \ref{betaPriorDist} and \ref{EPriorDist} in Eq. \ref{factorizedJointDistEq}, enables us to compute the joint probability. The edges of network are generated based on the value of variables $G$ and $B$ and form the network. \\
As described in \cite{yang2012structure}, in general real world networks have dense overlapping areas, because nodes in overlapping areas are members of more than one community. Therefore, the possibility of edge creation in overlapping areas is more than non-overlapping areas. In our model, according to Eq. \ref{edgeProbEq}, the two nodes with similar community memberships have more chance to be 
connected. Therefore, the generated networks will have dense overlapping areas.
 The proof is as follow:
\begin{equation}
\begin{split}
p(e_{ij}|\beta,\alpha) = \sum_{r=1}^{K}p(e_{ij},g_{ij}=r|\beta,\alpha) \\ =  \sum_{r=1}^{K}p(e_{ij}|g_{ij}=r,\beta,\alpha) p(g_{ij}=r|\alpha) \\ = \sum_{r=1}^{K} p(g_{ij}=r|\alpha) \beta_{ir}\beta_{jr}
\end{split}
\label{edgeProbEq}
\end{equation}
From Eq. \ref{edgeProbEq}, it can be seen that as much as two nodes ($i$,$j$) share more common communities, they have more chance to be connected, which means that overlapping nodes (members of more than one community), have more chance to be connected and overlapping areas will be dense parts of network.

For the inference section, for simplicity, only existing edges are considered. Because our goal is to find important nodes to determine communities, only nodes with high importance (high $\beta$) should be found. According to our model, an edge $e_{ij}$ doesn't form when two nodes $i$ and $j$ are unimportant in a community, so discarding unformed edges has no remarkable effect on the result \cite{ren2009simple}. The probability of network given other parameters is changed as follow:
\begin{equation}
\begin{split}
P(A|G,B,\alpha,\gamma) = \prod_{i=1}^{N} \prod_{j\in Neigh(i)} \beta_{i,g_{ij}}\beta_{j,g_{ij}}
\end{split}
\label{edgeDist2}
\end{equation}
where $Neigh(i)$ is the set of neighbors of node $i$. 
\subsection{Extension to Dynamic Networks}
In this part, we extend the generative model to dynamic networks. Consider a network at $T$ different time-steps called \textit{snapshot}. Nodes and edges can be added or removed during time periods. The variables at time $t$ will be shown having  $t$ as superscript, e.g. $G^t, A^t$ and $B^t$.  Dynamic methods are offline or online. In offline approach, at every snapshot, all past and future snapshots of the network are available, but in online methods only access to previous time snapshots is available. We chose online approach for our method, because in real world, we don't have knowledge about the future and online approach is more appropriate for real data.
\par
\begin{figure}[!t]
\centering
\includegraphics[width=0.36\textwidth]{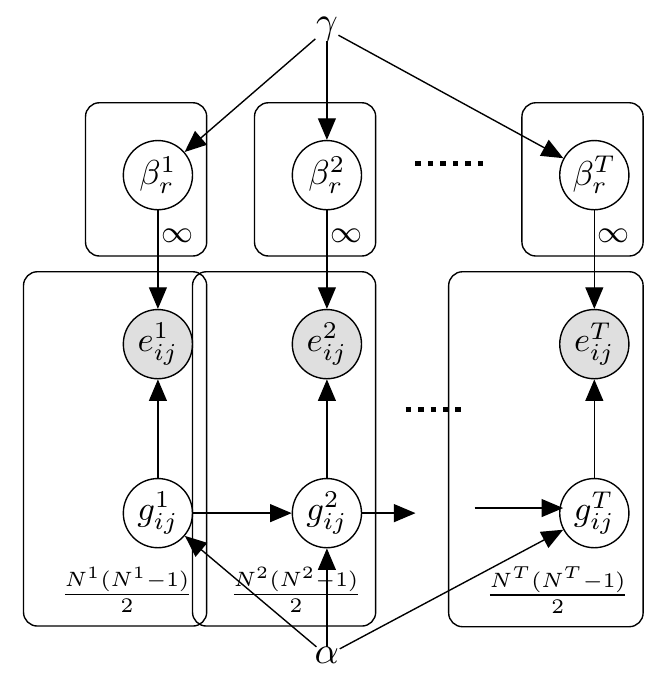}
\caption{Extended graphical model for dynamic network}
\label{graphicalModel2Fig}
\end{figure}

 To extend the static model, we use Recurrent Chinese Restaurant Process (RCRP) instead of CRP with some modifications. In the typical RCRP model, the value of parameters of an existing table at time $t$ depends on their value at time $t-1$, but in our model, we discard this limitation to have a flexible model. We consider that memberships are consistent over time and the degree of memberships can vary in time snapshots, so we discard the dependency between $\beta$s in snapshots and assume that only the community memberships in each time snapshot depend on the community memberships in the previous time snapshot. Moreover, we consider that all the history of communities is summarized in the last snapshot. The proposed graphical model for dynamic network is depicted in Fig. \ref{graphicalModel2Fig}.
 According to the graphical model, the joint probability of variables is as follows:
\begin{flalign}
& p(G^{1:T},B^{1:T} ,A^{1:T},\alpha ,\gamma ) & \nonumber\\ 
&=  \prod_{t=1}^T p(A^t|B^t ,G^t)  \prod_{t=2}^T p( G^t|G^{t-1}, \alpha) p(G^1|\alpha)  \prod_{t=1}^T p(B^t|\gamma)& 
\label{jointDynaDist}
\end{flalign}
\section{Inference and algorithm}
In this section, we use inference methods to find the parameters of the generative model, described in Section. \ref{method:sec}. We divide this section into two subsection. In the first part, we use Gibbs sampling method to find parameters of the model in static network and then extend the solution for dynamic networks and propose final algorithm which is our goal. This division makes the method steps easier to understand.
\subsection{Static Model Inference}
Our model utilizes a posterior inference on $G$ and $B$ to find the community memberships. Because the exact inference in CRP-based models is intractable \cite{neal2000markov}, we use approximate inference methods. One category of methods for approximating inference is the Markov Chain Monte Carlo (MCMC) methods. Gibbs sampling is one of the most popular MCMC methods, which we will use to solve this problem. In this approximation method, samples are drawn from the conditional distribution of each variable given current value of the other variables. 
In this section, we will provide the sampling equations for variables $G$ and $B$.\\
\textbf{Sampling $\mathbf{\beta_{r}}$}: The conditional distribution of $\beta_r$ given other variables has a closed form and will be a Dirichlet distribution. We first calculate the conditional distribution of $\beta_{ir}$ as follows:
\begin{equation}
\begin{split}
p(\beta _{ir}|G,A,\beta _{\backslash {{\beta }_{ir}}},\alpha ,\gamma)\propto \prod_{j=1}^N \beta _{ir}^{{e_{ij}}I[g_{ij}=r]} \beta _{ir}^{\gamma_{i}-1} \\ =\beta _{ir}^ {\sum_{j=1}^N e_{ij}I[g_{ij}=r ]}\beta _{ir}^{\gamma -1}
\label{condBeta1Eq}
\end{split}
\end{equation}
where $I$ is the indicator function and $\beta _{\backslash {{\beta }_{ir}}}$ means all variables in $\beta$ except $\beta_{ir}$. We define new variable $N_{ir} = \sum_{j=1}^N e_{ij}I[g_{ij}=r]$, that shows the number of connected edges to node $i$ which belong to community $r$. Thus Eq. \ref{condBeta1Eq} can be rewritten as:
\begin{equation}
p(\beta _{ir}|G,A,\beta _{\backslash {{\beta }_{ir}}},\alpha ,\gamma) \propto \beta _{ir}^{{{N}_{ir}}+\gamma-1}
\end{equation}
And we can conclude that:
\begin{equation}
{\beta }_{r}|(G,A,{{\beta }_{\backslash {{\beta }_{r}}}},\alpha,\gamma) \sim Dir( N_r+\gamma )
\label{betaCondFinalEq}
\end{equation}
Dirichlet distribution is a known distribution and it is easy to generate samples from it. \\

\textbf{Sampling $\mathbf{G}$}: For sampling from variable $G$, we estimate conditional distribution of every $g_{ij}$ for existing edges:
\begin{equation}
\begin{split}
 p(g_{ij}|G_{\backslash {g_{ij}}},A,B ,\alpha ,\gamma ) \propto \sum_{r=1}^K n_r p(e_{ij}|\beta_{r}) I\left[g_{ij}=r\right] \\
+ \alpha p(e_{ij} | G_0) I\left[g_{ij}=K+1\right]
\end{split}
\label{condG1Eq}
\end{equation}
where, $n_r$ is the size of community $r$ and $G_0$ is the base distribution in CRP. In this case, $G_0$ is a Dirichlet distribution with hyper-parameter $\gamma$. 
For sampling from distribution of Eq. \ref{condG1Eq}, we need to compute the second term of it:
\begin{equation}
\begin{split}
p(e_{ij} | G_0) I_{[g_{ij}=K+1]} =\int_{\beta_{K+1}} p(e_{ij}|\beta_{K+1})p(\beta_{K+1}|\gamma)  \\
= \int_{\beta_{K+1}} \beta_{i(K+1)}\beta_{j(K+1)}\frac{1}{C(\gamma)} \prod_{l=1}^N \beta_{l(K+1)}^{\gamma_{l} -1}  \\
= \frac{C(\gamma_{new})}{C(\gamma_{K+1})} \int_{\beta_{K+1}} Dir(\gamma_{new}) = \frac{C(\gamma_{new})}{C(\gamma)}
\end{split}
\end{equation}
\begin{equation}
\gamma_{new} = [\gamma_{new(1)},\ldots ,\gamma_{new (N)}]
\end{equation}
\begin{equation}
\gamma_{new (l)} = \left\{ 
\begin{array}{l l}
    \gamma_{(l)} & \quad l\neq i,j\\
&     \\
    \gamma_{(l)}+1 & \quad  l = i,j\\
\end{array} \right.
\end{equation}
So Eq. \ref{condG1Eq} can be rewritten as below:
\begin{equation}
p(g_{ij}=k|G_{\backslash {g_{ij}}},A, B ,\alpha ,\gamma ) \propto \left\{ 
\begin{array}{l l}
    n_k \beta_{ik}\beta_{jk} & \quad \scalebox{0.9}{existing group}\\
&     \\
    \alpha \frac{C(\gamma_{new})}{C(\gamma)}  & \quad \scalebox{0.9}{new group}\\
\end{array} \right.
\label{GCondFinalEq}
\end{equation}
\begin{equation}
\gamma_{new (l)} = \left\{ 
\begin{array}{l l}
    \gamma_{(l)} & \quad l\neq i,j\\
&     \\
    \gamma_{(l)}+1 & \quad  l = i,j\\
\end{array} \right.
\end{equation}
where, $C(\gamma)$ is the normalization constant of Dirichlet distribution.
Sampling from Eq. \ref{GCondFinalEq} can be easily achieved by using a random number generators. \\
\textbf{Determining communities from samples}: After $S$ repetition of sample generation from Eq. \ref{betaCondFinalEq} and Eq. \ref{GCondFinalEq}, we have $S$ samples of variables $B$ and $G$. To infer the community memberships of nodes, in each round of sampling process, we define the variable $u_r = \frac{n_r}{M}\beta_r$ ($M$ is the number of edges) and then for each node $i$, we divide the membership degree of node $i$ in community $s$  ($u_{is}$) to the maximum value of membership degree of node $i$ in all communities. If the result is more than a threshold $\theta$, node $i$ will be assigned to community $s$. This helps us to avoid insignificant communities, and be able to detect overlapping and non-overlapping communities. Now for each round of sampling, the network is divided into a set of communities. To choose the best community structure among samples, we use overlapping modularity and select the sample with maximum modularity \cite{shen2009quantifying,xie2013overlapping}.
\subsection{Dynamic Model Inference}
Because of independence assumption between $B^t$ and $B^{t-1}$ in the model, inference on variable $B$ in the dynamic model remains unchanged and is the same as static model (Eq. \ref{betaCondFinalEq}), but the rest is changed as follows. \\
\textbf{Inference on $\mathbf{G^t}$}: Inference on variable $G^t$ at time $(t=1)$ is as Eq. \ref{GCondFinalEq}, and for time  ($t\neq 1$) is as follows:
 \begin{align}
& p(g^t_{ij}=k|G^t_{\backslash {g^t_{ij}}},A^t,B^t ,\alpha ,\gamma ) \propto & \nonumber \\
& (n_{k,t}^{(ij)}+n_{k,t-1}) \beta_{ik}^t\beta_{jk}^t  I \left[ k\in G^{t-1}\right] & \nonumber \\
& +  (n_{k,t}^{(ij)}) \beta_{ik}^t\beta_{jk}^t I \left[ k\in G^t \; and\; k\notin G^{t-1}\right] & \nonumber\\ 
& +  \alpha \frac{C(\gamma_{new})}{C(\gamma)} I \left[(k\notin G^{t-1}\; and\; k\notin G^t \right] & \label{GDynaCondFinalEq}
\end{align}
where, $I$ is the indicator function and $C(\gamma)$ is the normalization constant of Dirichlet distribution.\\ 
\textbf{Initialization and Algorithm}: We initialize variables in $G$ randomly for the first time snapshot ($t=1$). At a time ($t\neq1$), for initialization, existing edges are assigned to the same group as the previous time snapshot and new ones are assigned randomly. This helps us to find consistent groups in a few number of sampling rounds. Moreover, to have faster convergence,
we initialize parameters $\beta$ to maximize the likelihood of the network. The likelihood of network is as follow:
\begin{equation}
\begin{aligned}
LL = P(A^t|G^t,\beta^t, \alpha, \gamma) &= \prod_{i=1}^N\prod_{j\in Neigh(i)} p({e}_{ij}^t|{g}_{ij}^t,\beta_{g_{ij}^t}^t) \\
 &= \prod_{i=1}^{N} \prod_{j\in Neigh(i)} \beta_{i,g_{ij}^t}^t\beta_{j,g_{ij}^t}^t
 \label{MaximizingLikelihood}
 \end{aligned}
\end{equation}
where $Neigh(i)$ is the set of neighbors of node $i$. To maximize Eq. \ref{MaximizingLikelihood}, we can write lagrange form of log(LL) with the constraint $\sum_{i=1}^N \beta_{ir}=1$:
\begin{equation}
\begin{split}
L = \sum_{i=1}^N \sum_{r=1}^K N_{ir}^t log(\beta_{ir}^t) +  \sum_{r=1}^K C_r (\sum_{i=1}^N \beta_{ir}^t-1)
\end{split}
\end{equation}
where $C_r$ is lagrange multiplier and $N_{ir}^t = \sum_{j=1}^{N^t} e_{ij}^t I[g_{ij}^t=r]$. If we differentiate $L$ with respect to $\beta{ir}$ and set it to zero, we will have $\beta_{ir}^t = \frac{N_{ir}^t}{\sum_{i=1}^{N^t} N_{ir}^t}$.
The pseudo code of DBOCD is presented in Alg. \ref{alg:DBOCD}.
\\\textbf{Time Complexity}: The complexity of generating a Dirichlet vector of dimension $N$ is $O(N)$. Hence, sampling $g_{ij}$s and $\beta_r$s in each round of sampling, has a time complexity $O(M+NK)$, where $M$ is the number of edges, $N$ is the number of nodes and $K$ is the number of communities. The total number of sampling rounds is a constant and can be ignored. Therefore, the time complexity of our model for each snapshot of the network is $O(M+NK_{max})$, where $K_{max}$ is the maximum number of communities over sampling rounds.
\begin{figure}[!ht]
\caption{DBOCD Algorithm for detecting communities over time} \label{alg:DBOCD}
\begin{algorithmic}[1]
\REQUIRE Set of Adjacency matrices over T snapshot(s), The membership threshold for node assignment in a group ($\theta$)\\
\ENSURE Founded Communities in Each Snapshot
\FOR{t=1 \TO T} 
		\IF{t=1} 
			\STATE{Initialize $G^1$ randomly}
			\STATE{Initialize $\beta_{ir}^1= \frac{N_{ir}^1}{\sum_{i=1}^N N_{ir}^1}$}
			\FOR{n=1 \TO number of needed samples}
				\STATE{Sample each $g_{ij}^1$ from \ref{GCondFinalEq}}
				\STATE{Sample each $\beta_{r}^1$ from \ref{betaCondFinalEq}}
				\STATE{$u_r^1 = \frac{n_r^1}{M^1}\beta_r^1$}
				\FOR{each node i}
					\STATE{$r = argmax_s \lbrace u_{is}^1, s = 1,\ldots , K^1 \rbrace$}
					\STATE{Assign node $i$ with $\frac{u_{is}^1}{u_{ir}^1}$ more than $\theta$ to community $s$}				
				\ENDFOR
			\ENDFOR
		\ELSE 
			\STATE{Initialize community membership of existing edges with founded communities at (t-1) and new ones randomly (Initializing $G^t$)} 
			\STATE{Initialize $\beta_{ir}^t = \frac{N_{ir}^t}{\sum_{i=1}^N N_{ir}^t}$}
			\FOR{n=1 \TO number of needed samples}
				\STATE{Sample each $g_{ij}^t$ from \ref{GDynaCondFinalEq}}
				\STATE{Sample each $\beta_{r}^t$ from \ref{betaCondFinalEq}}
				\STATE{$u_r^t = \frac{n_r^t}{M^t}\beta_r^t$}
				\FOR{each node i}
					\STATE{$r = argmax_s \lbrace u_{is}^t, s = 1,\ldots , K^t \rbrace$}
					\STATE{Assign node $i$ with $\frac{u_{is}^t}{u_{ir}^t}$ more than $\theta$ to community $s$}				
				\ENDFOR
			\ENDFOR
		\ENDIF
	\STATE{Select best sample with maximum modularity}
\ENDFOR
\end{algorithmic}
\end{figure}
\section{Experiments}
\label{experiment:sec}
In this section, we first describe the way we select the best value for hyper-parameters, then we describe our synthetic and real datasets, and finally, the results of DBOCD on synthetic and real datasets are provided. Since the main goal of this paper is introducing a dynamic method for community detection for dynamic networks, we don't provide the results of comparing BOCD (the static proposed method) with other static works.
\subsection{Hyper-parameter Selection}
\label{hyperpar:subsec}
\begin{figure*}[!htb]
\centering
\includegraphics[width=1\textwidth]{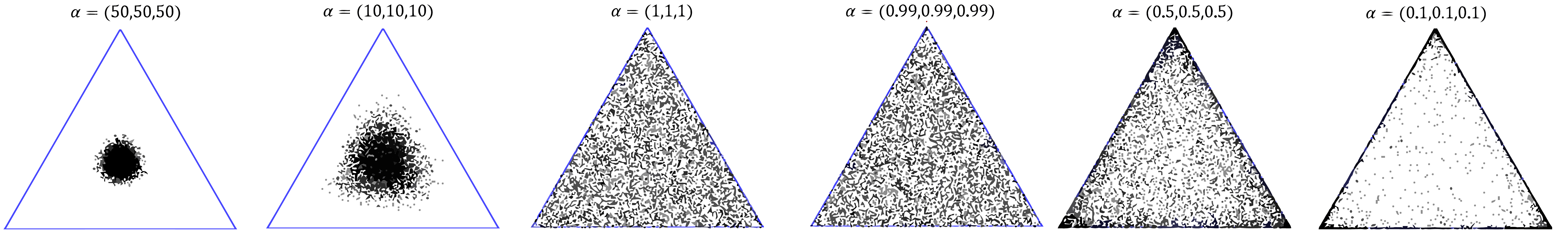}
\caption{Dirichlet distribution over 3-event probability simplex with different hyperparameters}
\label{dirichletSimplexFig}
\end{figure*}

In the proposed method, $\gamma$ and $\alpha$ are hyper-parameters. $\gamma$ is a vector of size $N$ which is the hyper-parameter of prior Dirichlet distribution on the importance of nodes in communities. Since there is no prior information about the network communities, we set all the entries in vector $\gamma$ to an identical value. In Dirichlet distribution, if we set all the entries of the hyper-parameter $\gamma$ to a value greater than one, a sample vector with equal values will have more chance to be generated. If we set all entries to one, the probability of all possible outcomes will be equal, and finally if the entries are set to a value less than one, the generated sample vector more probably will have a few large elements. 
This is shown in Fig. \ref{dirichletSimplexFig} that illustrates dirichlet distribution over 3-event probability simplex. 

We assume that there are a few important persons in each community in real world (e.g. admin, manager), and  set all elements of $\gamma$ to 0.1. Therefore, it is more probable that the generated $\beta_r$ be a vector with a few important nodes in community $r$.
\par
\begin{figure*}[!t]
    \centering
        \subfloat[Birth-Death]{\includegraphics[width=0.32\textwidth]{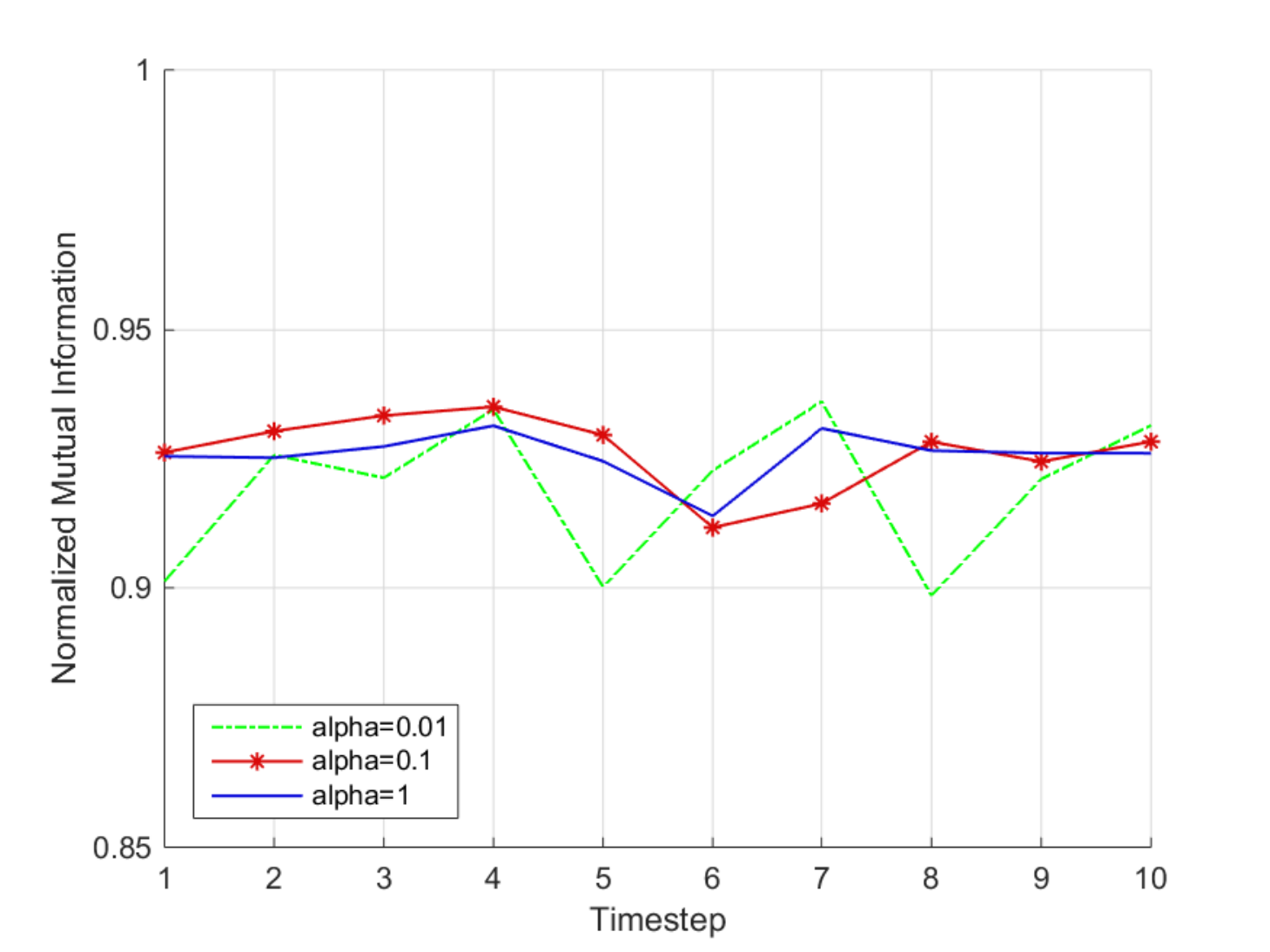}
        \label{fig:birth-death_alpha}}
    \hfill
        \subfloat[Expansion-Contraction]{\includegraphics[width=0.32\textwidth]{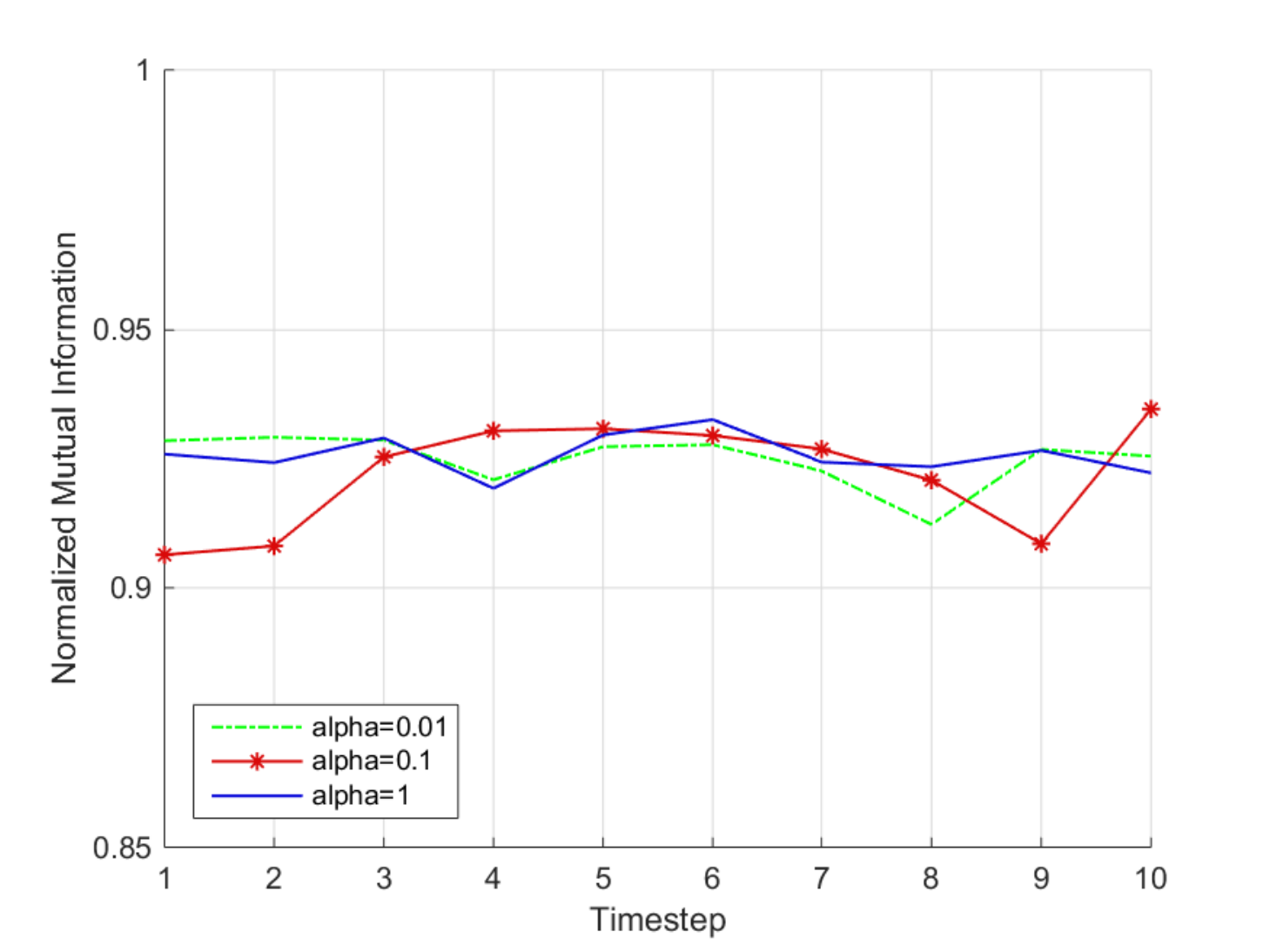}
        \label{fig:expand-contract_alpha}}
        \hfill
        \subfloat[Merging-Splitting]{\includegraphics[width=0.32\textwidth]{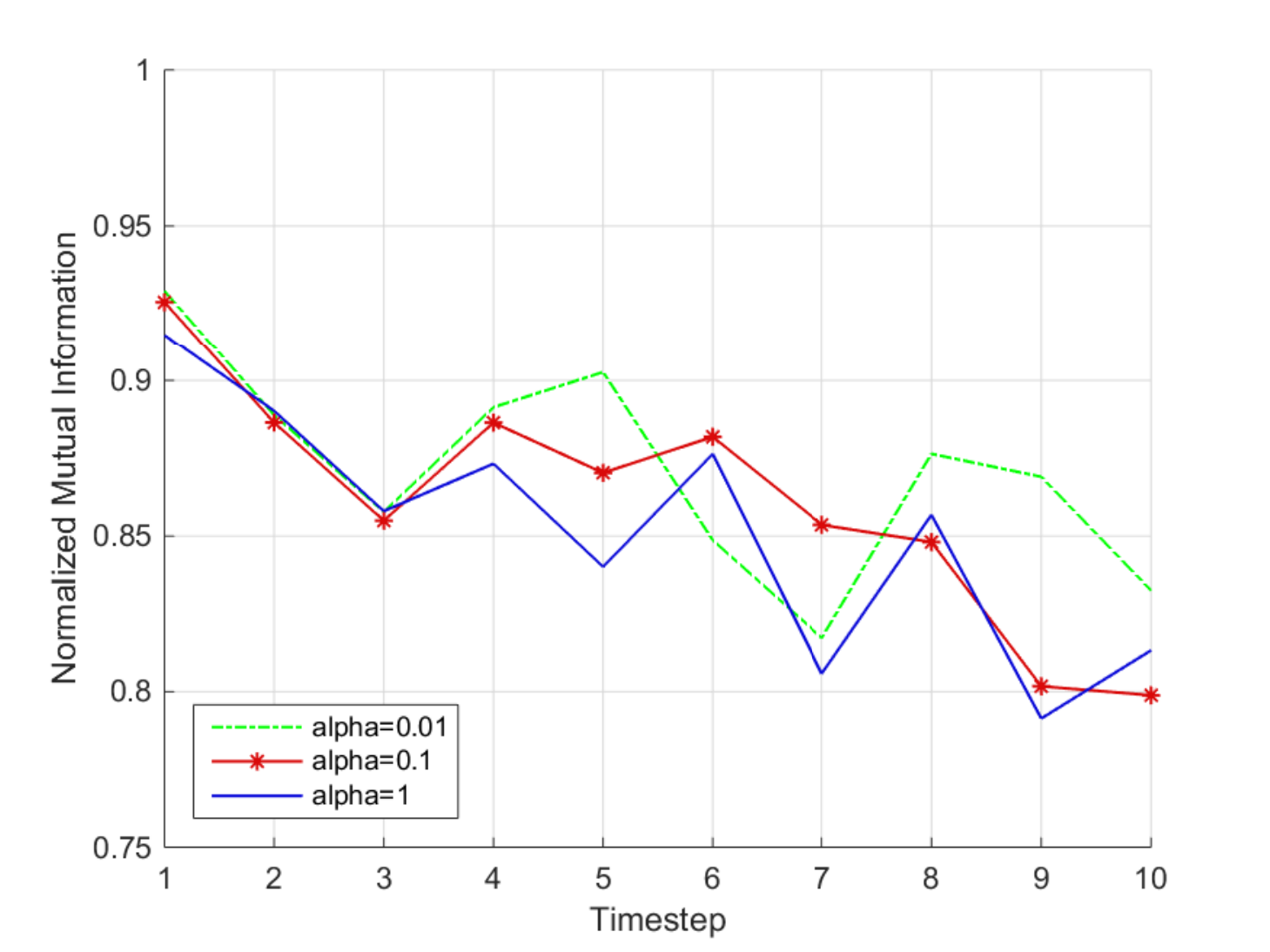}
        \label{fig:merge-split_alpha}}
    \caption{Results of different values of $\alpha$ on evolving datasets}
    \label{alphaChangeFig}
\end{figure*}

 The probability of selecting a new community in our method is proportional to hyper-parameter $\alpha$. 
To select the best value for hyper-parameter $\alpha$, we ran DBOCD with different values of $\alpha$ on a synthetic dataset generated by a dynamic benchmark proposed in \cite{greene2010tracking} .Greene's benchmark is a dynamic extension of the static LFR benchmark \cite{lancichinetti2009benchmarks} and models different types of evolutions of communities over time. In this benchmark, at each snapshot, community memberships are assigned to nodes and then the network is constructed based on the setting of dataset, so the consistency of networks is preserved over time. We have generated 3 dynamic networks. Concurrent birth-death, expansion-contraction and merging-splitting of communities over time snapshots are modeled in 3 different datasets. Common setting of datasets is reported in Table \ref{alphaSetting:tab}. 
\begin{table}[!t]
\renewcommand{\arraystretch}{1.3}
\caption{Common setting of all synthetic datasets}
\label{alphaSetting:tab}
\centering
\begin{tabular}{|c|c|c|c|c|c|c|}
\hline
N & Avg Deg & Max Deg & On & Om & $\mu$ & T \\
\hline
1000 & 40 & 60 & 40 & 4 & 0.3 & 10 \\
\hline
\end{tabular}
\end{table}
In the table, N is the number of nodes, (on) shows the number of overlapping nodes, (om) shows the number of communities that every overlapping node belong to, $\mu$ shows the mixing parameter that control the community structure of network and T is the number of snapshots. In all datasets, $10\%$ of nodes change their community memberships randomly relative to the previous snapshot. Because we have ground-truth of memberships, Normalized Mutual Information (NMI) \cite{lancichinetti2009detecting} is used as a measure for comparing true communities with the ones extracted by DBOCD. NMI is a number between 0 and 1 that evaluates similarity between two sets of communities, and has its maximum value when the similarity is maximum. In the first snapshot, 100 samples are produced and in the other snapshots 50 samples are generated by DBOCD. 
The results of experiments are shown in Fig. \ref{alphaChangeFig}. As illustrated, the effect of the hyper-parameter $\alpha$ is small in every snapshot (the values differ by less than $0.1$). Since in the first snapshot, we initialize the number of communities with $\frac{N}{5}$, the value of $\alpha$ is not critical, and the number of communities decreases in the generation process. In the other snapshots, because of considering the previous snapshot and stability of communities over time, the effect of parameter $\alpha$ is negligible. In the proposed method, we choose $\alpha = 0.1$ as the default value, because for this value the results are more stable than the other cases.
\subsection{Experiments Settings}
We chose FacetNet \cite{lin2008facetnet}, DSBM \cite{yang2011detecting} and AFFECT \cite{xu2014adaptive} as competitors to compare the proposed method with them. DSBM and AFFECT cannot detect overlapping communities, but FacetNet can detect overlapping ones. The threshold for extracting communities from soft memberships is set to $0.7$. In FacetNet, parameter $\alpha$ is set to 0.1 which produced the best results. To do inference in DBOCD, 100 samples are generated for the first snapshot and 50 samples for other snapshots. A "DBOCD-With Modularity" label, shows the sample which  has maximum modularity. A "DBOCD-Best Sample" label, shows the sample with maximum NMI. In DSBM, we set all the parameters to default values. For the AFFECT algorithm, we ran all 3 clustering methods that proposed in the original paper and selected the one with the maximum average NMI over time. FacetNet needs the number of communities, we initialize it with the true number. DSBM has the assumption that the number of communities remains unchanged over time. We initialize it with the true number of communities in the first snapshot.  In the AFFECT method, the number of communities is extracted with silhouettes width \cite{rousseeuw1987silhouettes}. Since all the methods are probabilistic ones, we ran each method 5 times and report the average results.
\begin{figure*}[!t]
    \centering
        \subfloat[Birth-Death]{\includegraphics[width=0.32\textwidth]{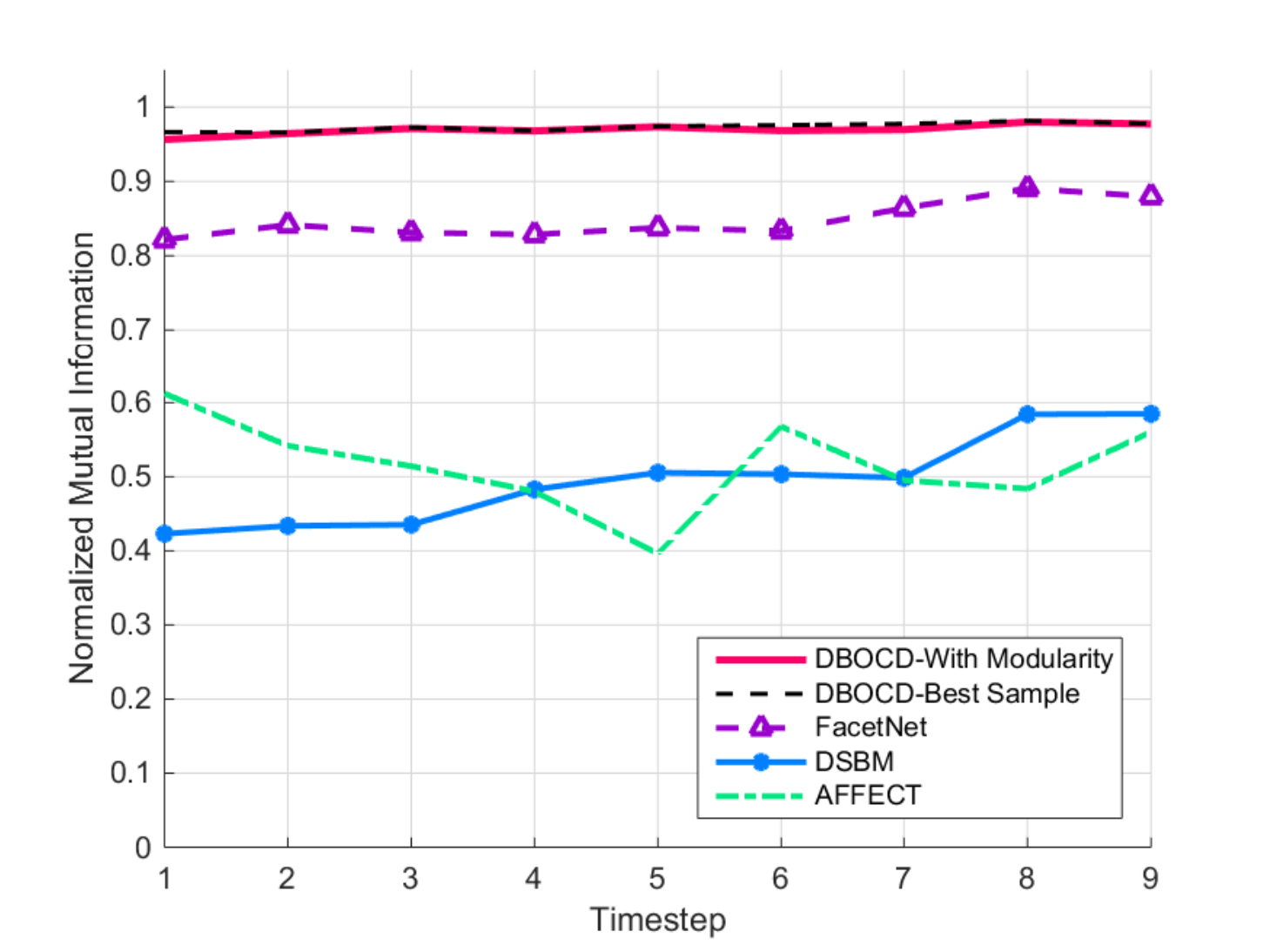}
        \label{fig:birth-death}}
        \hfill
        \subfloat[Expansion-Contraction]{\includegraphics[width=0.32\textwidth]{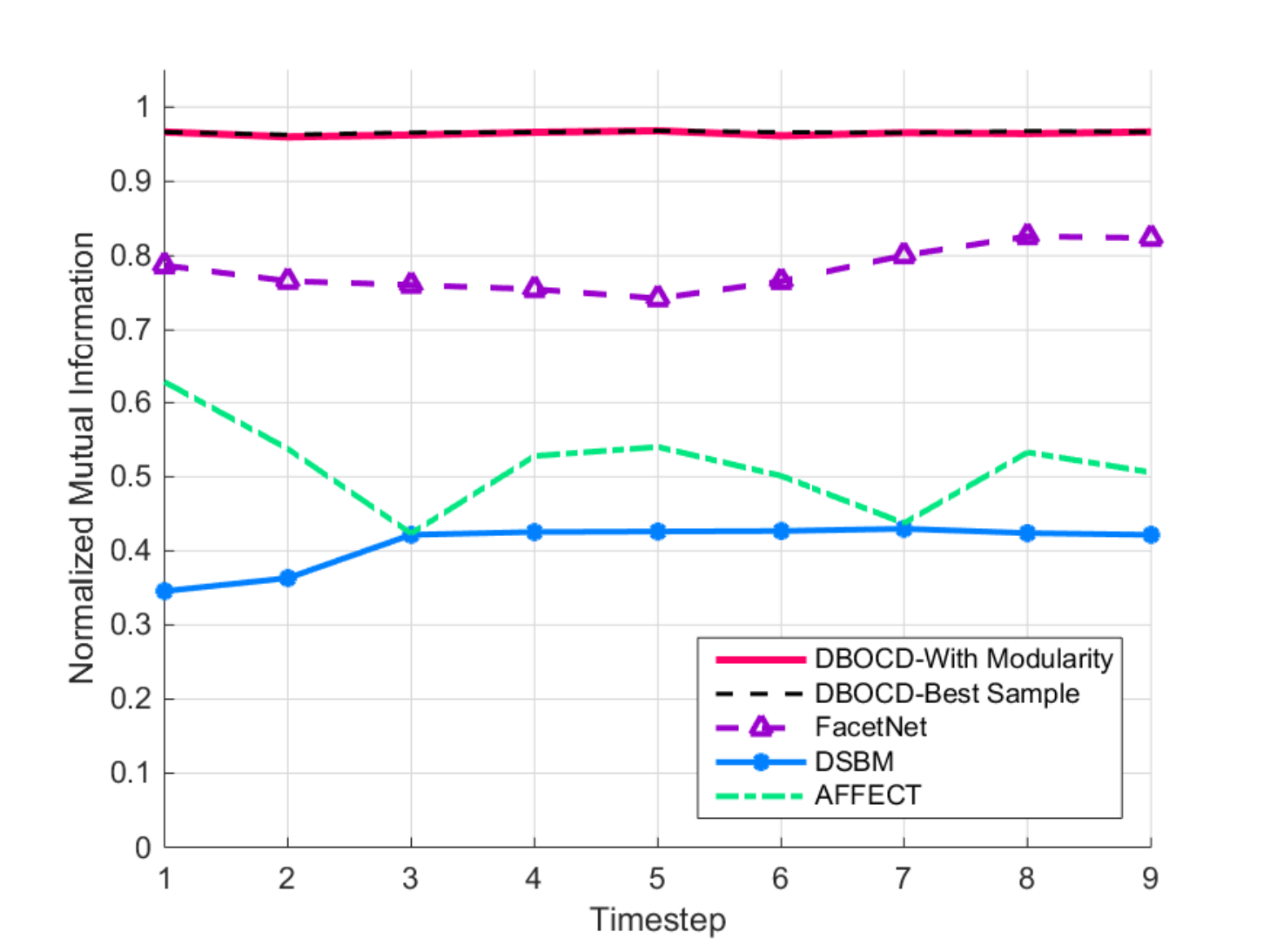}
        \label{fig:expand-contract}}
        \hfill
        \subfloat[Merging-Splitting]{\includegraphics[width=0.32\textwidth]{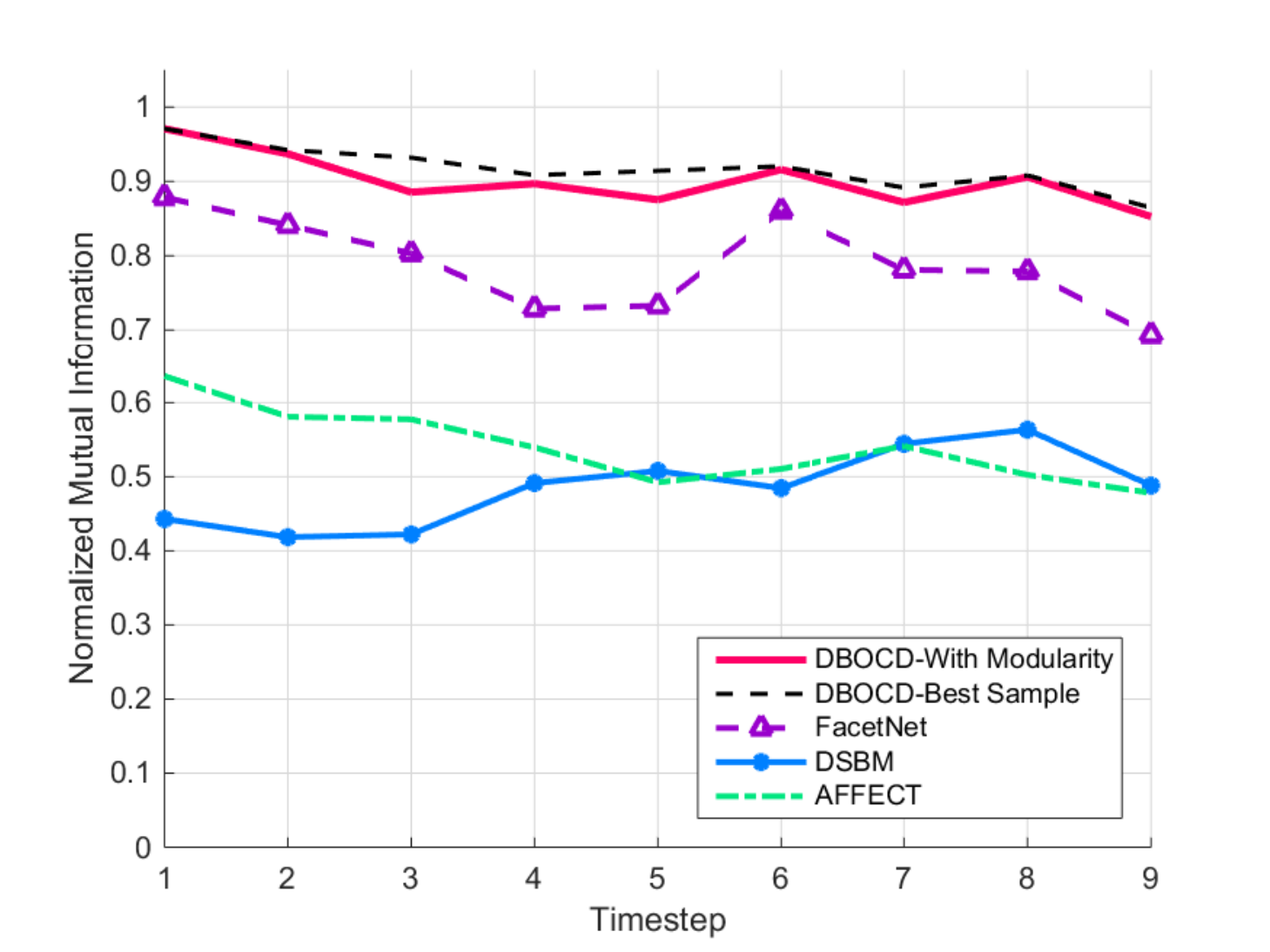}
        \label{fig:merge-split}}
    \caption{Results of methods on different types of synthetic data}
    \label{fig:syntheticResult}
\end{figure*} 

\subsection{Synthetic Results}
We generated the synthetic data by using a dynamic benchmark proposed in \cite{greene2010tracking}. The benchmark is introduced in Section \ref{hyperpar:subsec}. We generated 3 dynamic networks with different types of community evolution over time. The generated datasets model concurrent birth-death, expansion-contraction and merging-splitting of communities over time snapshots. Common setting of all datasets is reported in Table \ref{commonSetting:tab}.
\begin{table}[!t]
\renewcommand{\arraystretch}{1.3}
\caption{Common setting of all synthetic datasets}
\label{commonSetting:tab}
\centering
\begin{tabular}{|c|c|c|c|c|c|c|}
\hline
N & Avg Deg & Max Deg & On & Om & $\mu$ & T \\
\hline
500 & 30 & 50 & 20 & 3 & 0.2 & 9 \\
\hline
\end{tabular}
\end{table}
In all datasets, $10\%$ of nodes change their community memberships randomly.  Generated datasets are different in types of evolution and the number of evolving communities. Since the generated data has ground truth communities, the NMI measure  is used to compare  the performance of different algorithms. 

The NMI results are illustrated in Fig. \ref{fig:syntheticResult}.
As we can see, DBOCD outperforms other methods and has stable results over time. We can also see that the modularity is a good measure to select the best sample and the results of the sample with maximum modularity is very close to the results for the sample with maximum NMI. Both AFFECT and DSBM have lower accuracy, because they cannot detect overlapping areas. FacetNet also has acceptable and stable performance, but, in merging-splitting, FacetNet has descending NMI over time. The unstable results on merging-splitting is because of more changes in communities in merge-split dataset and FacetNet is not able to handle it.
\par
\begin{figure*}[!htb]
    \centering
        \subfloat[Birth]{\includegraphics[width=0.32\textwidth]{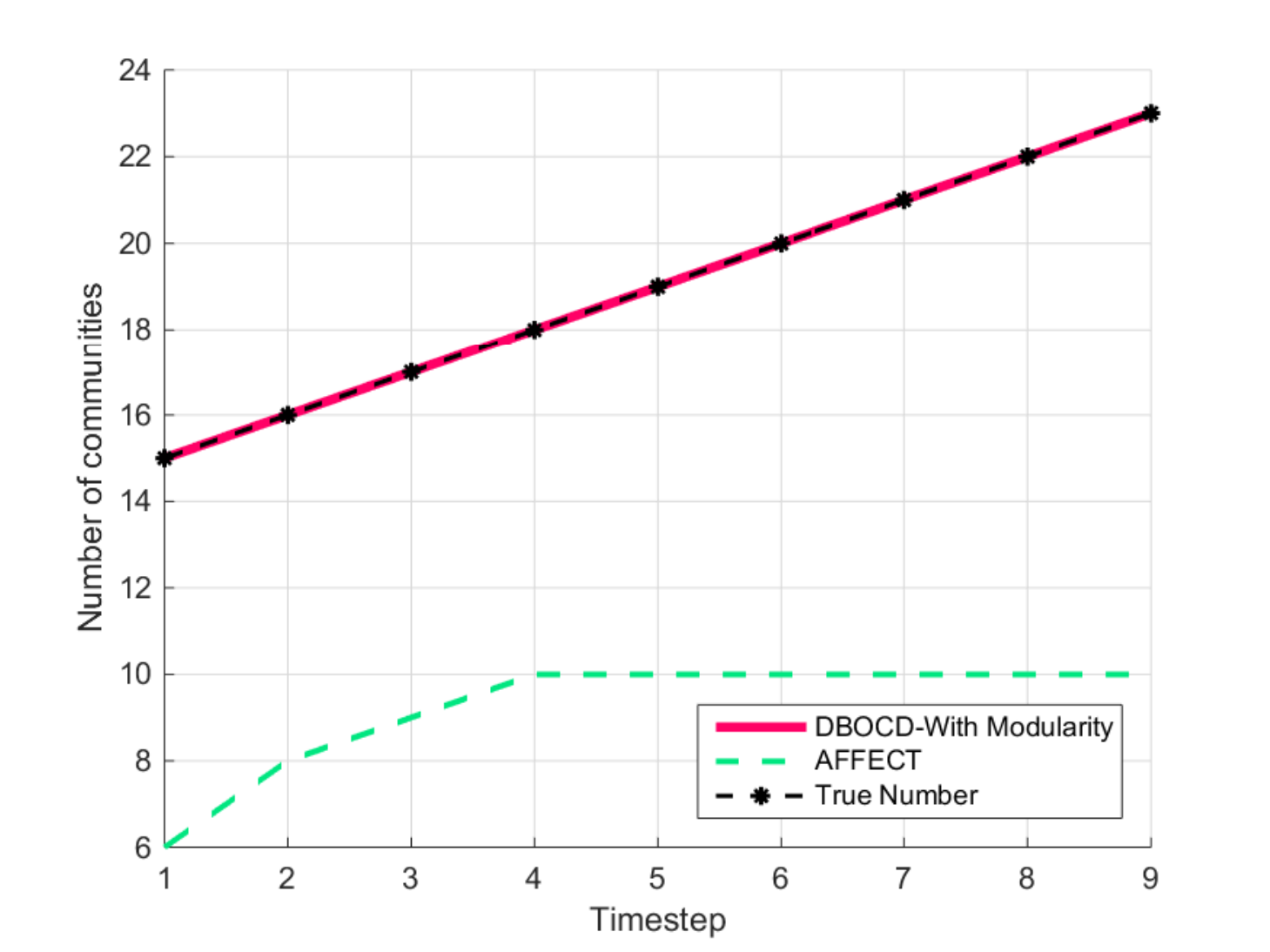}
        \label{fig:birth_K}}
        \hfill
        \subfloat[Death]{\includegraphics[width=0.32\textwidth]{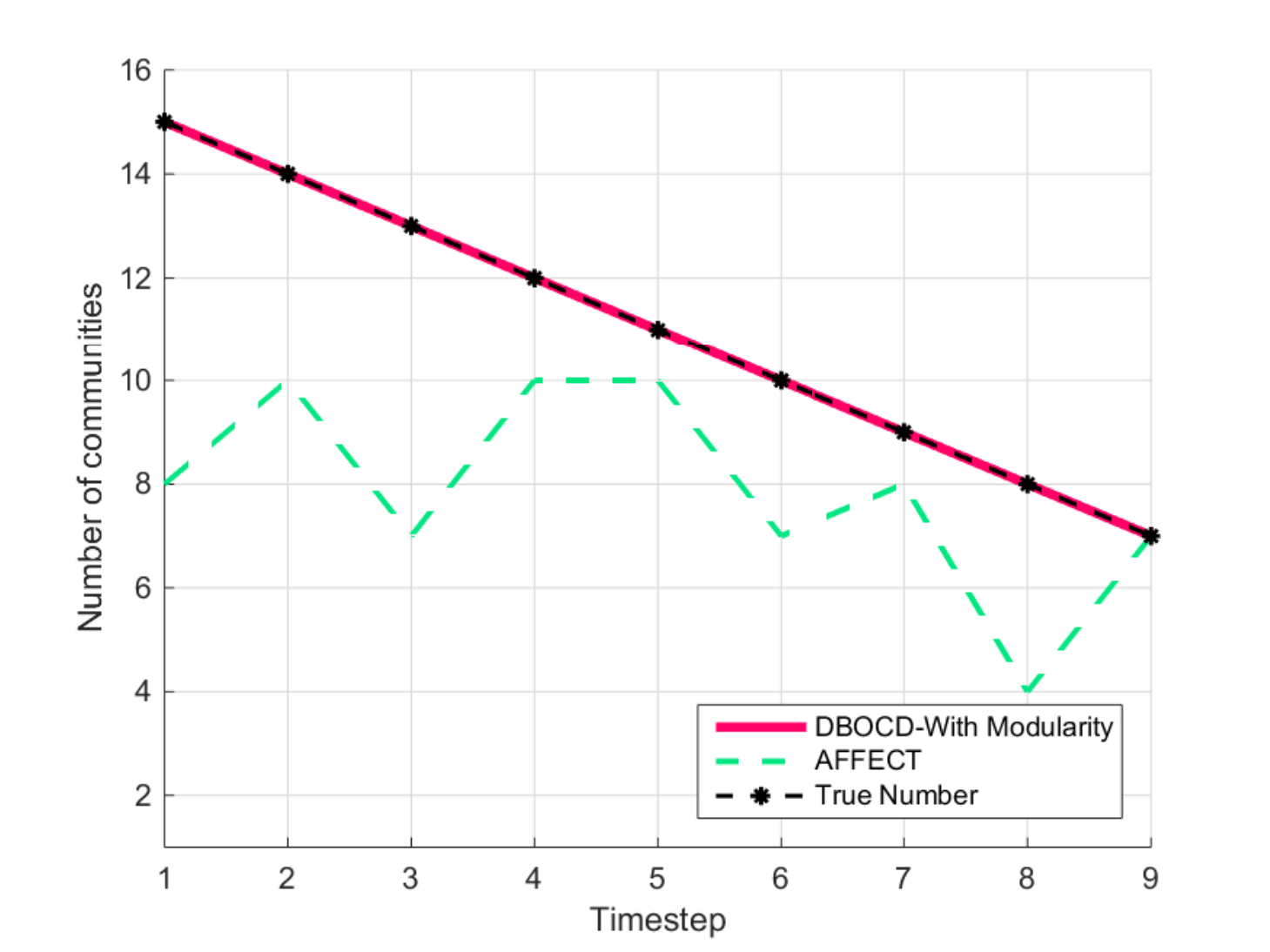}
        \label{fig:death_K}}
    \hfill
        \subfloat[Birth-Death]{\includegraphics[width=0.32\textwidth]{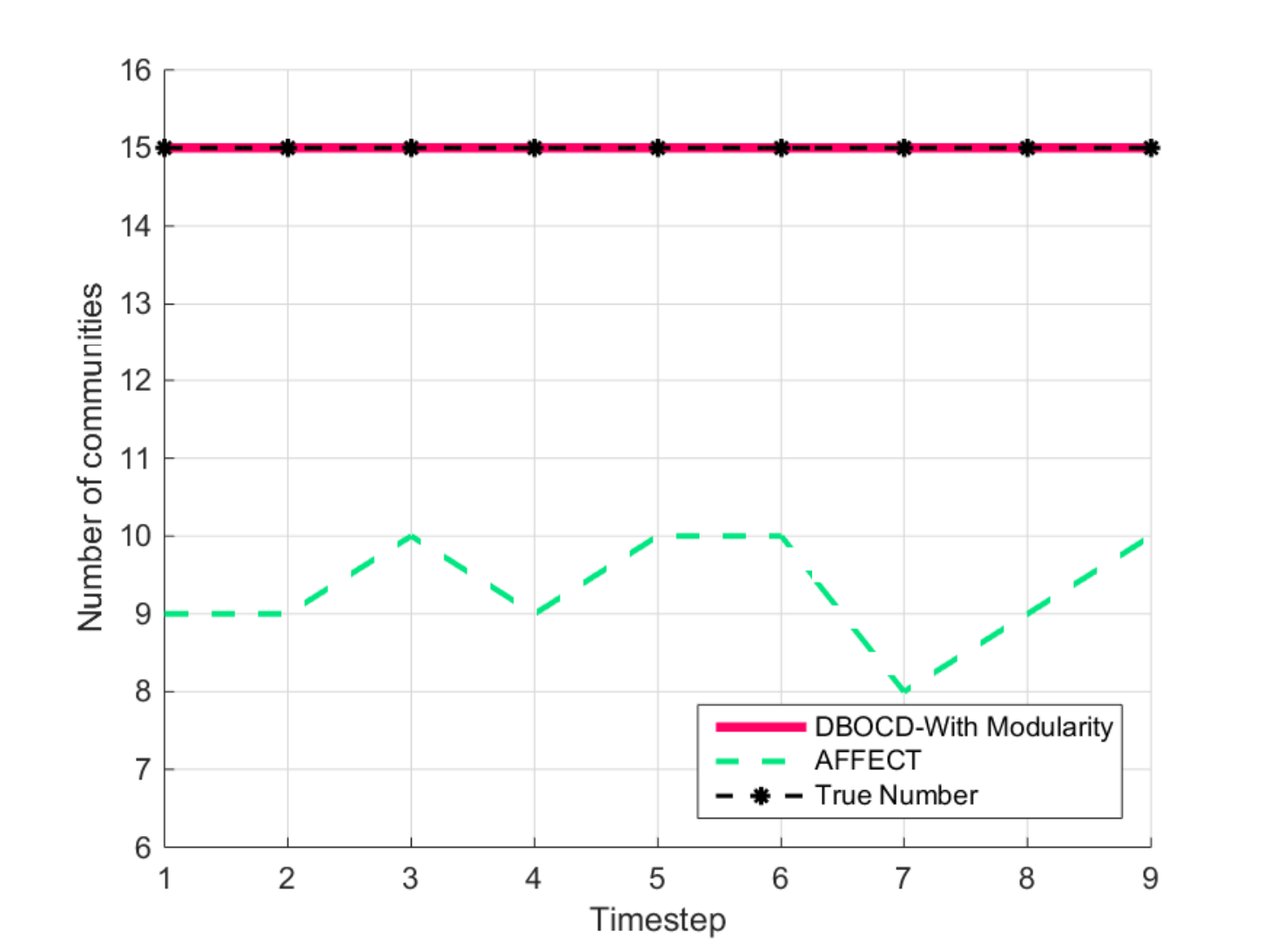}
        \label{fig:birth-death_K}}
\\
        \subfloat[Split]{\includegraphics[width=0.32\textwidth]{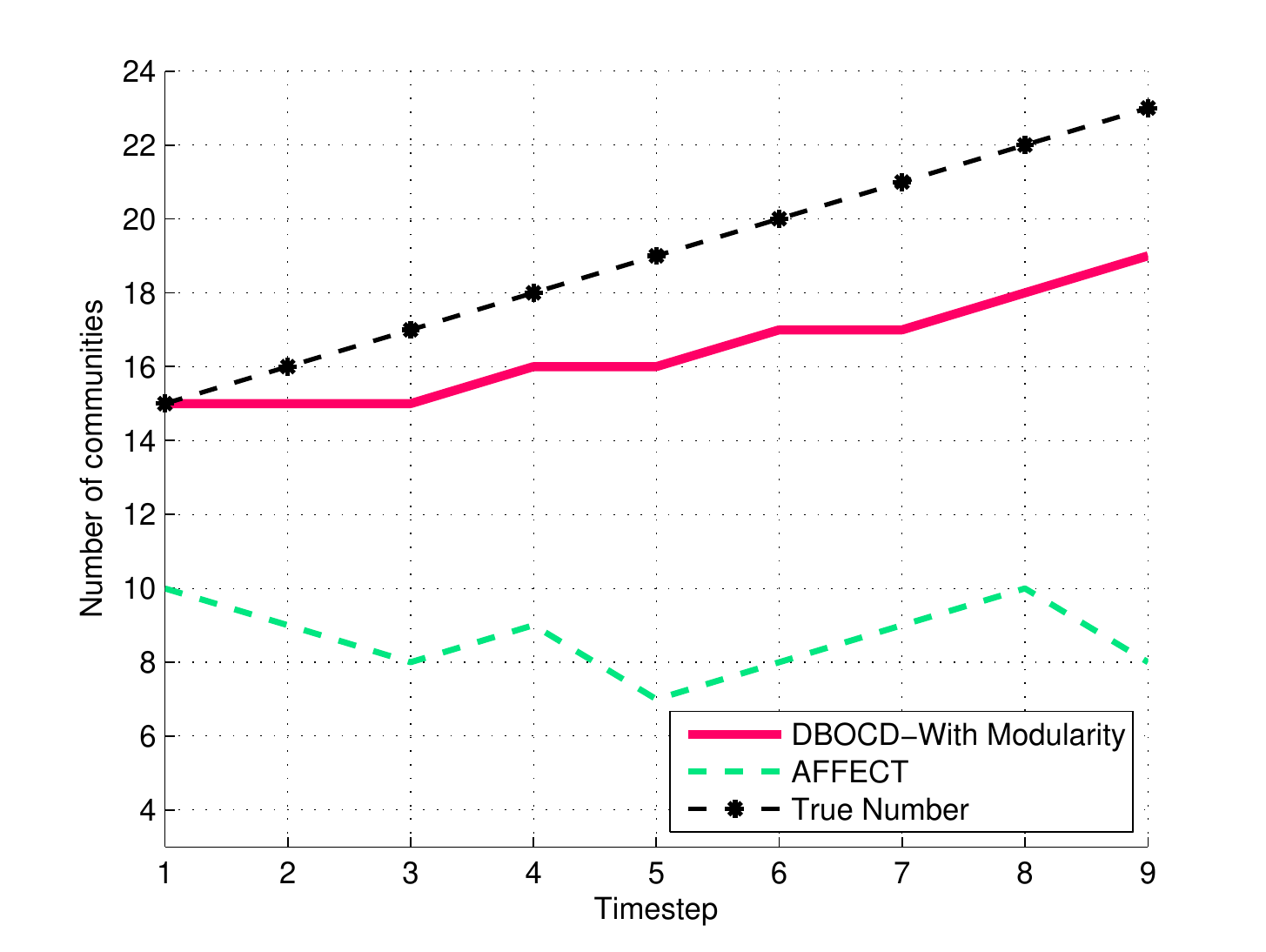}
        \label{fig:split_K}}
	\hfill
        \subfloat[Merge]{\includegraphics[width=0.32\textwidth]{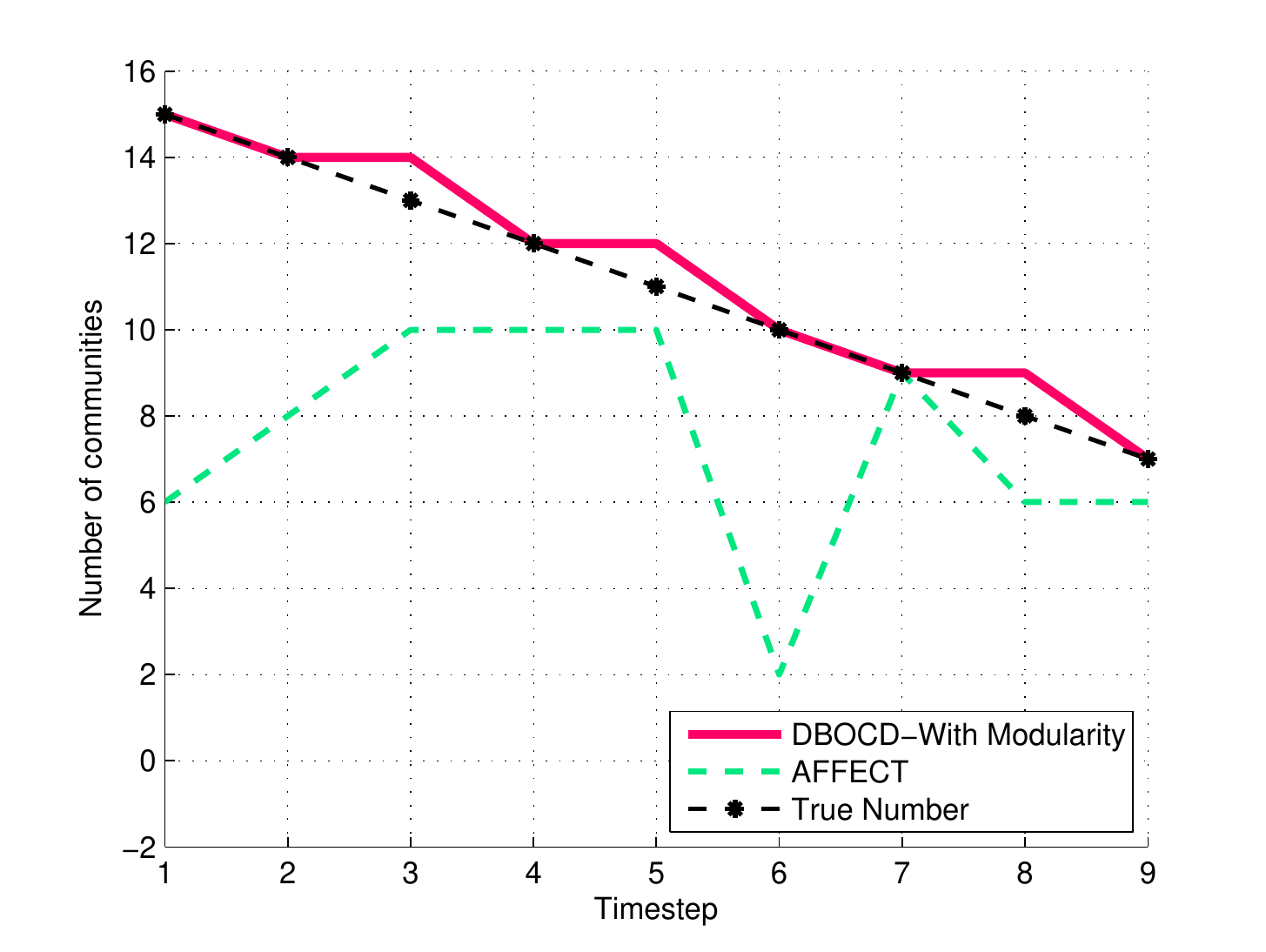}
        \label{fig:merge_K}}
        \hfill
        \subfloat[Merge-Split]{\includegraphics[width=0.32\textwidth]{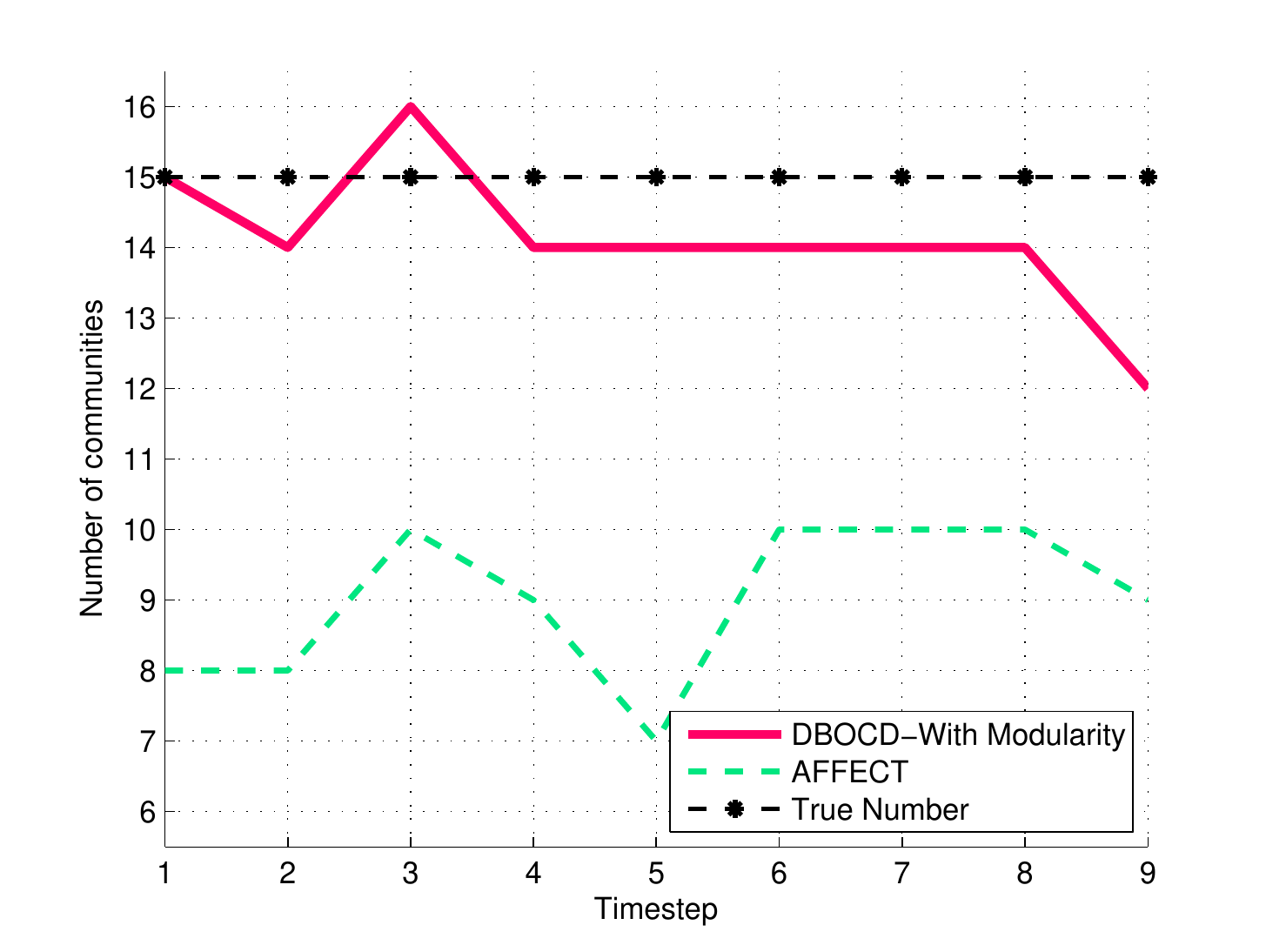}
        \label{fig:merge-split_K}}
        \caption{Comparison between number of communities detected by DBOCD and AFFECT}
    \label{fig:syntheticResult_K}
\end{figure*}

 We also compared the number of communities which is extracted by DBOCD and AFFECT algorithms on birth, death, concurrent birth-death, split, merge and concurrent merge-split datasets. In the birth process, members of new community are chosen randomly from all other communities, but in the split process a specific community is chosen randomly and split to two smaller communities. In the death process a community is destroyed and its members are distributed among other groups, but in the merge process two communities are chosen randomly and are joined together. In the birth-death and the merge-split, creation and ruin occur simultaneously, therefore the number of communities remains unchanged. Results are shown in Fig. \ref{fig:syntheticResult_K}. It can be seen that DBOCD can capture the exact changes in number of communities in birth, death and birth-death. The process of alteration in the number of communities in merge, split and merge-split. AFFECT cannot even find the gradient of changes in the number of communities in both cases. In split process, a small community is created, and in merge process a large community is formed. In both processes, the change in network is abruptive. This is the cause that DBOCD cannot find the exact number of communities.

\subsection{Real Dataset Results}
\begin{figure}[!ht]
    \centering
        \subfloat[NEC Nodes]{\includegraphics[width=0.23\textwidth]{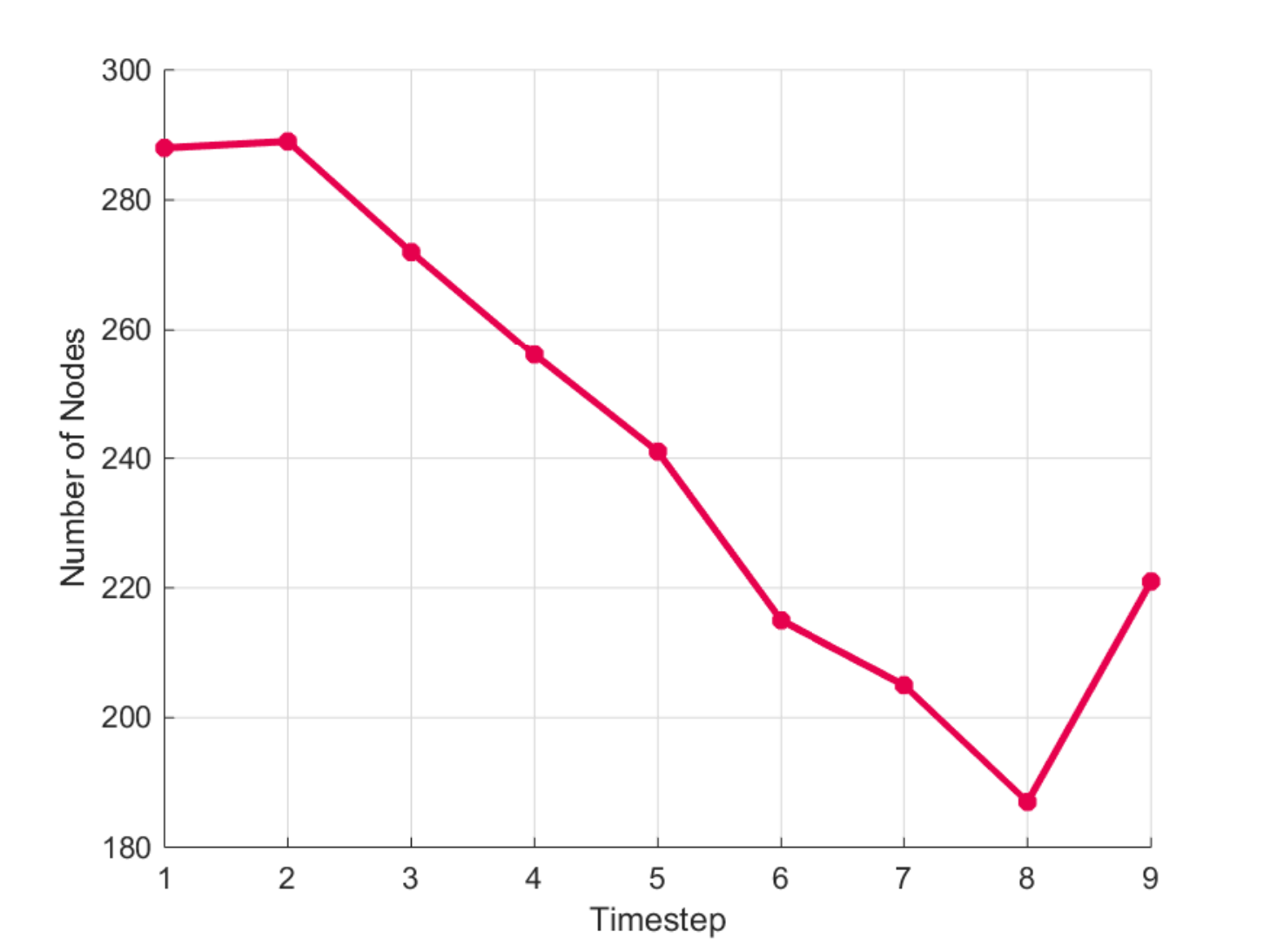}
        \label{fig:NEC_node}}
        \hfill
        \subfloat[NEC Edges]{\includegraphics[width=0.23\textwidth]{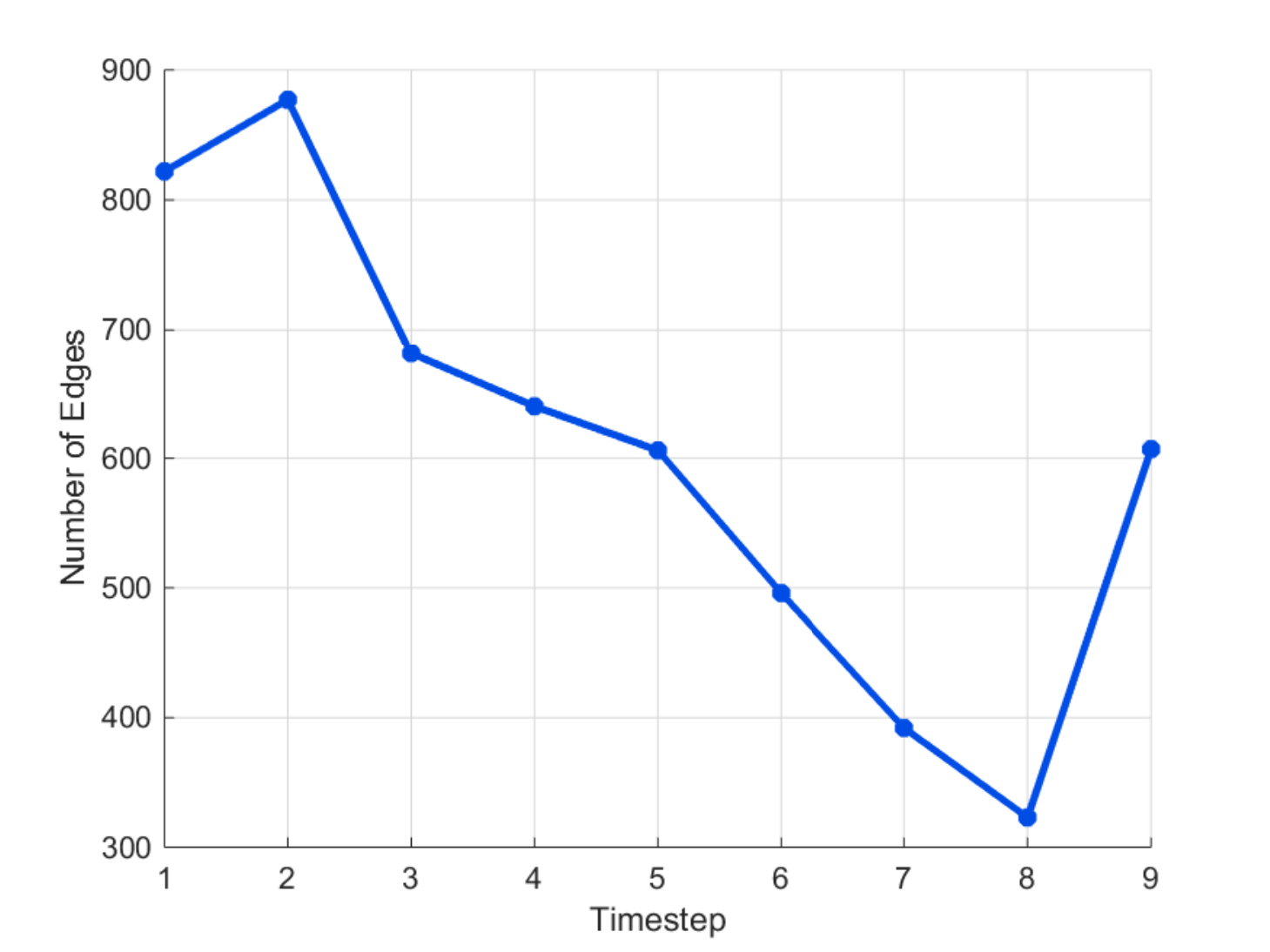}
        \label{fig:NEC_edge}}
\\
        \subfloat[DBLP Nodes]{\includegraphics[width=0.23\textwidth]{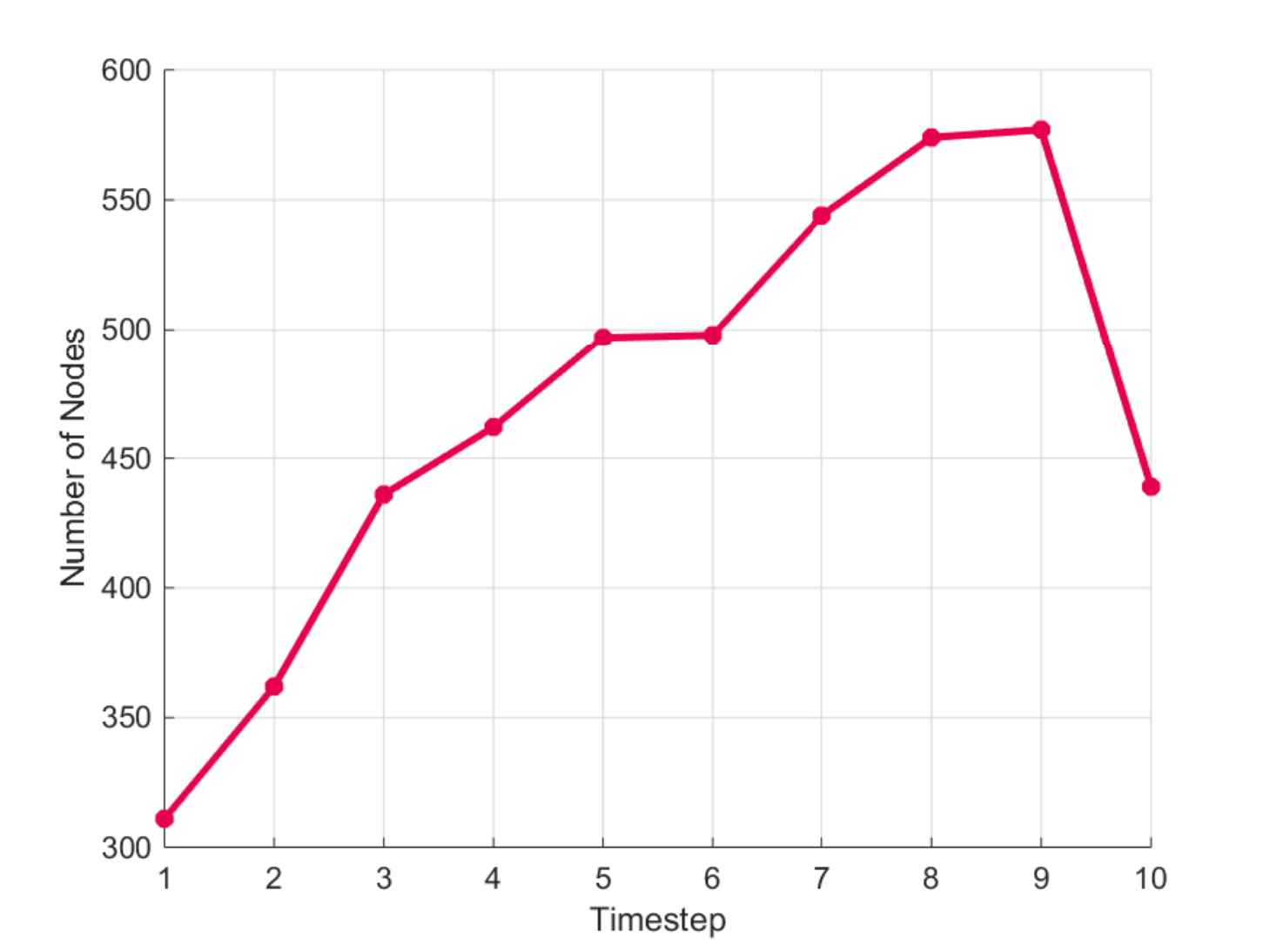}
        \label{fig:DBLP_node}}
        \hfill
        \subfloat[DBLP Edges]{\includegraphics[width=0.23\textwidth]{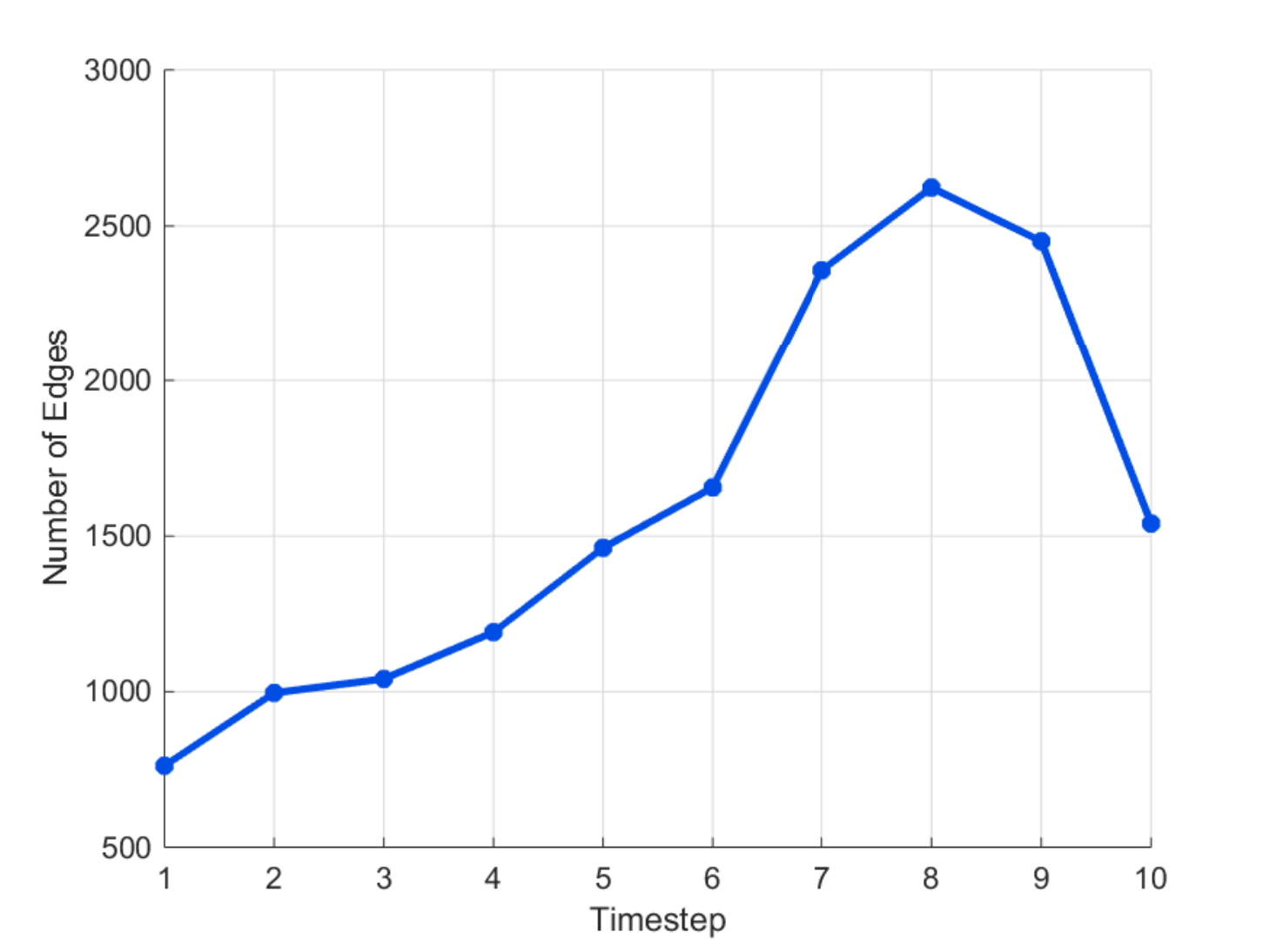}
        \label{fig:DBLP_edge}}
    \caption{Evolution of real datasets}
    \label{fig:real_data}
\end{figure}

We also examined DBOCD on two real datasets, NEC blog and DBLP paper co-authorship datasets. Both datasets are considered unweighted and undirected. NEC blog dataset is composed of 404 blogs and 148,681 links among them during 15 months. Because of abrupt diminishing in the numbers of links after the 10th month, nodes and links in period 10'th to 15'th months are aggregated into the 9th time step. DBLP dataset includes information about co-authorships in 28 conferences over 10 years (1997-2006). Nodes and edges evolution of both real datasets has been shown in Fig. \ref{fig:real_data}

The number of nodes and edges in NEC blog decrease over time, while in DBLP increase. Since there is no ground truth communities in real data, we use overlapping modularity \cite{shen2009quantifying,xie2013overlapping} for comparisons. The same as experiments on synthetic data, we compare DBOCD with FacetNet, AFFECT and DSBM. The parameters of methods have been set as before. Because we don't know the true number of communities, we test all the numbers in  the range [1-20] and select the value that has maximum value of average modularity over time for the methods which need the number of communities to be known. 

\begin{figure}[!ht]
	\centering
	        \includegraphics[width=0.35\textwidth]{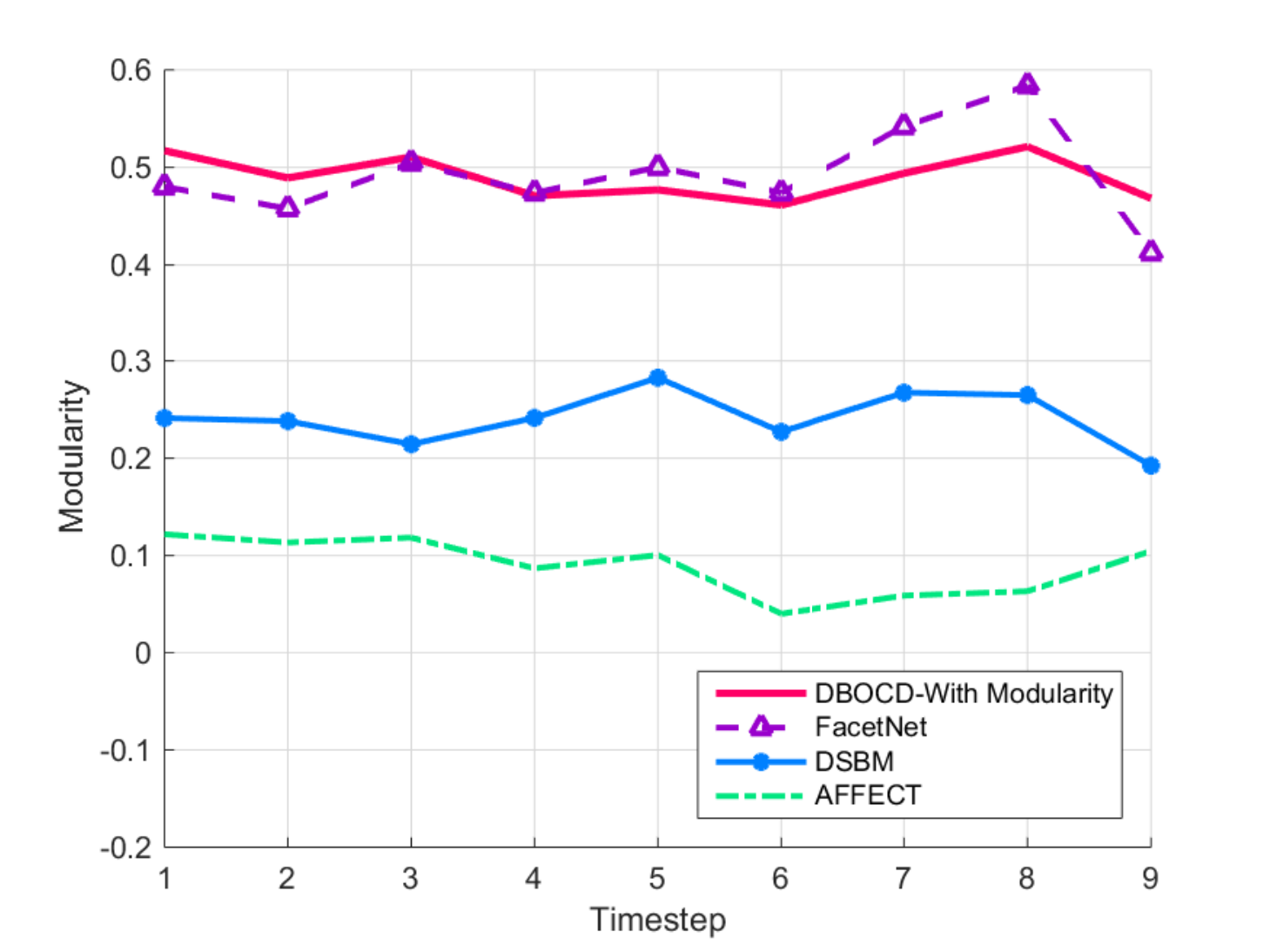}
	        \caption{Results on NEC dataset}
	        \label{fig:NEC_res}
	    \end{figure}

\begin{figure}[!ht]
	    \centering
	        \includegraphics[width=0.35\textwidth]{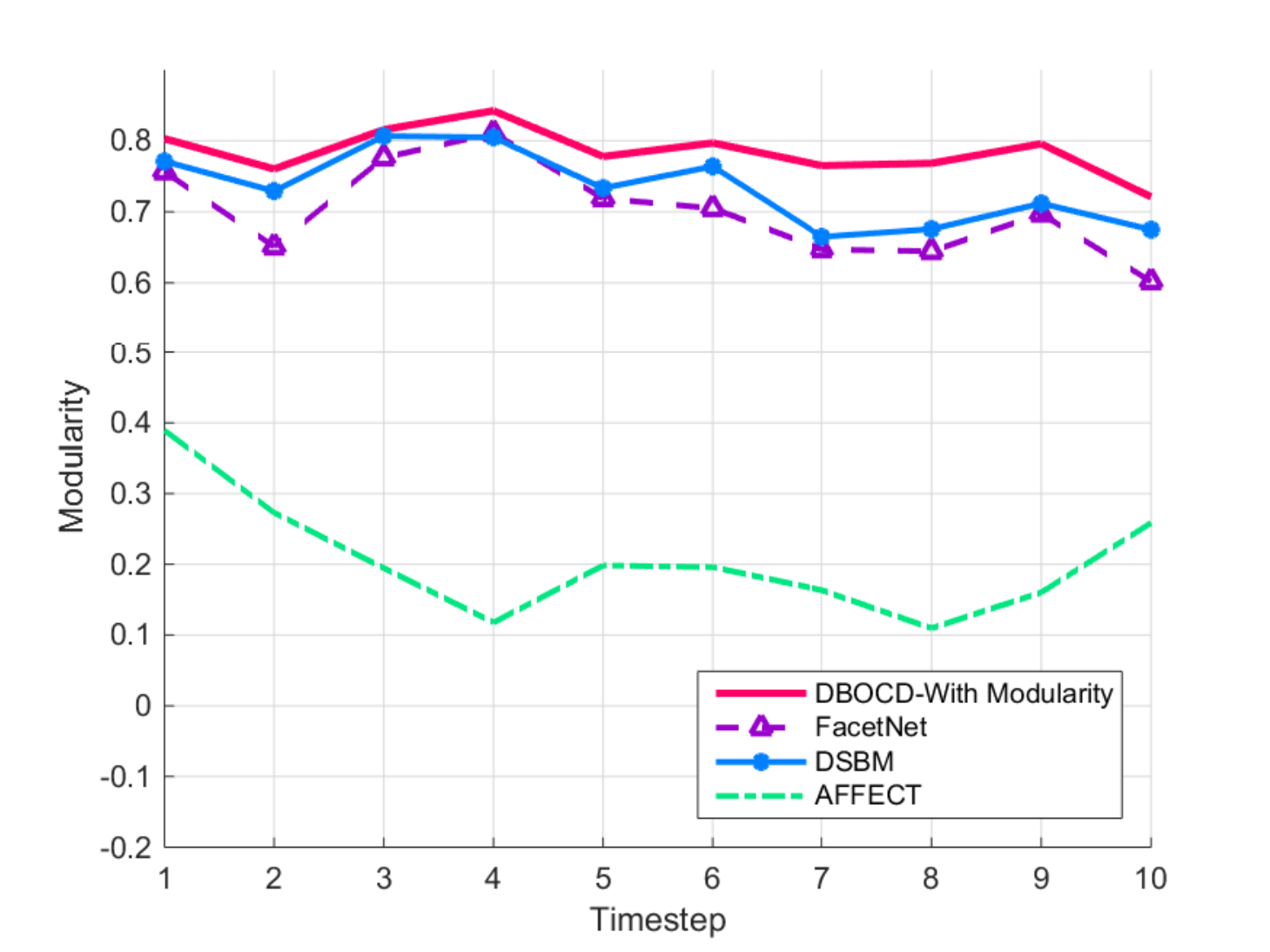}
	        \caption{Results on DBLP dataset}
	        \label{fig:DBLP_res}
   \end{figure}

The results on real datasets are illustrated in Fig. \ref{fig:NEC_res} and Fig. \ref{fig:DBLP_res}.
Although NEC blog is decreasing in the number of nodes and edges and DBLP is increasing, DBOCD has stable results on both of them, but other methods have many changes over time. Only in some snapshots of NEC blog, FacetNet has better results than DBOCD. In both datasets DBOCD has lower slope through time which can be interpreted that DBOCD detects consistent communities over time. Remarkable point is that DSBM as a non-overlapping detector has better results than FacetNet on DBLP dataset. It can be interpreted that DBLP is a non-overlapping dataset, and since DBOCD has the maximum modularity, it means DBOCD can detect both overlapping and non-overlapping communities with good performance.

\section{Conclusion}
\label{conclusion:sec}
In this paper, we presented a Dynamic Bayesian generative model to generate dynamic networks with overlapping community structure.  The model was named Bayesian Overlapping Community Detector (DBOCD). Because of using Recurrent Chinese Restaurant Process and link communities in each snapshot, the communities are detected in polynomial time complexity without any prior knowledge about the number of communities and also the stability of determined communities over time is preserved. DBOCD can find the soft memberships of nodes in communities in discrete time snapshots of network and handle the addition and deletion of nodes, edges and communities. The experimental results on the synthetic and real data showed that DBOCD can outperform other recent popular methods in different types of community evolution. 

In future work, we plan to apply the variational inference methods to infer the parameters of our model. This may also help to make the proposed method faster.

%
%
%
%
%
%
%
%
\ifCLASSOPTIONcaptionsoff
  \newpage
\fi



\bibliographystyle{IEEEtran}
\bibliography{Bibliography}

\begin{thebibliography}{10}
\providecommand{\url}[1]{#1}
\csname url@samestyle\endcsname
\providecommand{\newblock}{\relax}
\providecommand{\bibinfo}[2]{#2}
\providecommand{\BIBentrySTDinterwordspacing}{\spaceskip=0pt\relax}
\providecommand{\BIBentryALTinterwordstretchfactor}{4}
\providecommand{\BIBentryALTinterwordspacing}{\spaceskip=\fontdimen2\font plus
\BIBentryALTinterwordstretchfactor\fontdimen3\font minus
  \fontdimen4\font\relax}
\providecommand{\BIBforeignlanguage}[2]{{%
\expandafter\ifx\csname l@#1\endcsname\relax
\typeout{** WARNING: IEEEtran.bst: No hyphenation pattern has been}%
\typeout{** loaded for the language `#1'. Using the pattern for}%
\typeout{** the default language instead.}%
\else
\language=\csname l@#1\endcsname
\fi
#2}}
\providecommand{\BIBdecl}{\relax}
\BIBdecl

\bibitem{girvan2002community}
M.~Girvan and M.~E. Newman, ``Community structure in social and biological
  networks,'' \emph{Proceedings of the national academy of sciences}, vol.~99,
  no.~12, pp. 7821--7826, 2002.

\bibitem{sahebi2011community}
S.~Sahebi and W.~W. Cohen, ``Community-based recommendations: a solution to the
  cold start problem,'' in \emph{Workshop on Recommender Systems and the Social
  Web, RSWEB}, 2011.

\bibitem{li2014clustering}
F.~Li, J.~He, G.~Huang, Y.~Zhang, and Y.~Shi, ``A clustering-based link
  prediction method in social networks,'' \emph{Procedia Computer Science},
  vol.~29, pp. 432--442, 2014.

\bibitem{fortunato2010community}
S.~Fortunato, ``Community detection in graphs,'' \emph{Physics Reports}, vol.
  486, no.~3, pp. 75--174, 2010.

\bibitem{palla2007quantifying}
G.~Palla, A.-L. Barab{\'a}si, and T.~Vicsek, ``Quantifying social group
  evolution,'' \emph{Nature}, vol. 446, no. 7136, pp. 664--667, 2007.

\bibitem{xu2014adaptive}
K.~S. Xu, M.~Kliger, and A.~O. Hero~Iii, ``Adaptive evolutionary clustering,''
  \emph{Data Mining and Knowledge Discovery}, vol.~28, no.~2, pp. 304--336,
  2014.

\bibitem{yang2011detecting}
T.~Yang, Y.~Chi, S.~Zhu, Y.~Gong, and R.~Jin, ``Detecting communities and their
  evolutions in dynamic social networks—a bayesian approach,'' \emph{Machine
  learning}, vol.~82, no.~2, pp. 157--189, 2011.

\bibitem{lin2008facetnet}
\BIBentryALTinterwordspacing
Y.-R. Lin, Y.~Chi, S.~Zhu, H.~Sundaram, and B.~L. Tseng, ``Facetnet: A
  framework for analyzing communities and their evolutions in dynamic
  networks,'' in \emph{Proceedings of the 17th International Conference on
  World Wide Web}, ser. WWW '08.\hskip 1em plus 0.5em minus 0.4em\relax ACM,
  2008, pp. 685--694. [Online]. Available:
  \url{http://doi.acm.org/10.1145/1367497.1367590}
\BIBentrySTDinterwordspacing

\bibitem{xie2013overlapping}
J.~Xie, S.~Kelley, and B.~K. Szymanski, ``Overlapping community detection in
  networks: The state-of-the-art and comparative study,'' \emph{ACM Computing
  Surveys (csur)}, vol.~45, no.~4, p.~43, 2013.

\bibitem{reichardt2006statistical}
J.~Reichardt and S.~Bornholdt, ``Statistical mechanics of community
  detection,'' \emph{Physical Review E}, vol.~74, no.~1, p. 016110, 2006.

\bibitem{blondel2008fast}
V.~D. Blondel, J.-L. Guillaume, R.~Lambiotte, and E.~Lefebvre, ``Fast unfolding
  of communities in large networks,'' \emph{Journal of Statistical Mechanics:
  Theory and Experiment}, vol. 2008, no.~10, p. P10008, 2008.

\bibitem{baumes2005finding}
J.~Baumes, M.~K. Goldberg, M.~S. Krishnamoorthy, M.~Magdon-Ismail, and
  N.~Preston, ``Finding communities by clustering a graph into overlapping
  subgraphs.'' \emph{IADIS AC}, vol.~5, pp. 97--104, 2005.

\bibitem{lancichinetti2011finding}
A.~Lancichinetti, F.~Radicchi, J.~J. Ramasco, S.~Fortunato \emph{et~al.},
  ``Finding statistically significant communities in networks,'' \emph{PloS
  one}, vol.~6, no.~4, p. e18961, 2011.

\bibitem{adamcsek2006cfinder}
B.~Adamcsek, G.~Palla, I.~J. Farkas, I.~Der{\'e}nyi, and T.~Vicsek, ``Cfinder:
  locating cliques and overlapping modules in biological networks,''
  \emph{Bioinformatics}, vol.~22, no.~8, pp. 1021--1023, 2006.

\bibitem{kumpula2008sequential}
J.~M. Kumpula, M.~Kivel{\"a}, K.~Kaski, and J.~Saram{\"a}ki, ``Sequential
  algorithm for fast clique percolation,'' \emph{Physical Review E}, vol.~78,
  no.~2, p. 026109, 2008.

\bibitem{airoldi2009mixed}
E.~M. Airoldi, D.~M. Blei, S.~E. Fienberg, and E.~P. Xing, ``Mixed membership
  stochastic blockmodels,'' in \emph{Advances in Neural Information Processing
  Systems}, 2009, pp. 33--40.

\bibitem{psorakis2011overlapping}
I.~Psorakis, S.~Roberts, M.~Ebden, and B.~Sheldon, ``Overlapping community
  detection using bayesian non-negative matrix factorization,'' \emph{Physical
  Review E}, vol.~83, no.~6, p. 066114, 2011.

\bibitem{he2014link}
D.~He, D.~Jin, C.~Baquero, and D.~Liu, ``Link community detection using
  generative model and nonnegative matrix factorization,'' \emph{PloS one},
  vol.~9, no.~1, p. e86899, 2014.

\bibitem{ahn2010link}
Y.-Y. Ahn, J.~P. Bagrow, and S.~Lehmann, ``Link communities reveal multiscale
  complexity in networks,'' \emph{Nature}, vol. 466, no. 7307, pp. 761--764,
  2010.

\bibitem{ren2009simple}
W.~Ren, G.~Yan, X.~Liao, and L.~Xiao, ``Simple probabilistic algorithm for
  detecting community structure,'' \emph{Physical Review E}, vol.~79, no.~3, p.
  036111, 2009.

\bibitem{leinhardt1976local}
S.~Leinhardt, ``Local structure in social networks,'' \emph{Sociological
  methodology}, vol.~7, pp. 1--45, 1976.

\bibitem{tantipathananandh2011finding}
C.~Tantipathananandh and T.~Y. Berger-Wolf, ``Finding communities in dynamic
  social networks,'' in \emph{Data Mining (ICDM), 2011 IEEE 11th International
  Conference on}.\hskip 1em plus 0.5em minus 0.4em\relax IEEE, 2011, pp.
  1236--1241.

\bibitem{liu2015label}
K.~Liu, J.~Huang, H.~Sun, M.~Wan, Y.~Qi, and H.~Li, ``Label propagation based
  evolutionary clustering for detecting overlapping and non-overlapping
  communities in dynamic networks,'' \emph{Knowledge-Based Systems}, vol.~89,
  pp. 487--496, 2015.

\bibitem{morup2012bayesian}
M.~M{\o}rup and M.~N. Schmidt, ``Bayesian community detection,'' \emph{Neural
  computation}, vol.~24, no.~9, pp. 2434--2456, 2012.

\bibitem{whang2013stochastic}
J.~J. Whang, P.~Rai, and I.~S. Dhillon, ``Stochastic blockmodel with cluster
  overlap, relevance selection, and similarity-based smoothing,'' in \emph{Data
  Mining (ICDM), 2013 IEEE 13th International Conference on}.\hskip 1em plus
  0.5em minus 0.4em\relax IEEE, 2013, pp. 817--826.

\bibitem{tang2011dynamic}
X.~Tang and C.~C. Yang, ``Dynamic community detection with temporal dirichlet
  process,'' in \emph{Privacy, Security, Risk and Trust (PASSAT) and 2011 IEEE
  Third Inernational Conference on Social Computing (SocialCom), 2011 IEEE
  Third International Conference on}.\hskip 1em plus 0.5em minus 0.4em\relax
  IEEE, 2011, pp. 603--608.

\bibitem{pitman2002combinatorial}
J.~Pitman \emph{et~al.}, ``Combinatorial stochastic processes,'' Technical
  Report 621, Dept. Statistics, UC Berkeley, 2002. Lecture notes for St. Flour
  course, Tech. Rep., 2002.

\bibitem{blei2011distance}
D.~M. Blei and P.~I. Frazier, ``Distance dependent chinese restaurant
  processes,'' \emph{The Journal of Machine Learning Research}, vol.~12, pp.
  2461--2488, 2011.

\bibitem{ahmed2008dynamic}
A.~Ahmed and E.~P. Xing, ``Dynamic non-parametric mixture models and the
  recurrent chinese restaurant process: with applications to evolutionary
  clustering.'' in \emph{SDM}.\hskip 1em plus 0.5em minus 0.4em\relax SIAM,
  2008, pp. 219--230.

\bibitem{griffiths2011indian}
T.~L. Griffiths and Z.~Ghahramani, ``The indian buffet process: An introduction
  and review,'' \emph{The Journal of Machine Learning Research}, vol.~12, pp.
  1185--1224, 2011.

\bibitem{yang2012structure}
J.~Yang and J.~Leskovec, ``Structure and overlaps of communities in networks,''
  \emph{arXiv preprint arXiv:1205.6228}, 2012.

\bibitem{neal2000markov}
R.~M. Neal, ``Markov chain sampling methods for dirichlet process mixture
  models,'' \emph{Journal of computational and graphical statistics}, vol.~9,
  no.~2, pp. 249--265, 2000.

\bibitem{shen2009quantifying}
H.-W. Shen, X.-Q. Cheng, and J.-F. Guo, ``Quantifying and identifying the
  overlapping community structure in networks,'' \emph{Journal of Statistical
  Mechanics: Theory and Experiment}, vol. 2009, no.~07, p. P07042, 2009.

\bibitem{greene2010tracking}
D.~Greene, D.~Doyle, and P.~Cunningham, ``Tracking the evolution of communities
  in dynamic social networks,'' in \emph{Advances in social networks analysis
  and mining (ASONAM), 2010 international conference on}.\hskip 1em plus 0.5em
  minus 0.4em\relax IEEE, 2010, pp. 176--183.

\bibitem{lancichinetti2009benchmarks}
A.~Lancichinetti and S.~Fortunato, ``Benchmarks for testing community detection
  algorithms on directed and weighted graphs with overlapping communities,''
  \emph{Physical Review E}, vol.~80, no.~1, p. 016118, 2009.

\bibitem{lancichinetti2009detecting}
A.~Lancichinetti, S.~Fortunato, and J.~Kert{\'e}sz, ``Detecting the overlapping
  and hierarchical community structure in complex networks,'' \emph{New Journal
  of Physics}, vol.~11, no.~3, p. 033015, 2009.

\bibitem{rousseeuw1987silhouettes}
P.~J. Rousseeuw, ``Silhouettes: a graphical aid to the interpretation and
  validation of cluster analysis,'' \emph{Journal of computational and applied
  mathematics}, vol.~20, pp. 53--65, 1987.

\end{thebibliography}
%
%
%

%
%
\newpage
\begin{IEEEbiography}[{\includegraphics[width=1in,height=1.25in,clip,keepaspectratio]{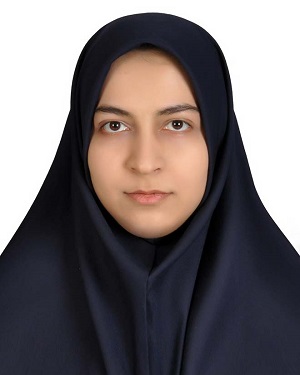}}]{Mahsa Ghorbani}
received her B.Sc. degree in software engineering and M.Sc. degree in artificial intelligence from Sharif University of Technology, Tehran, Iran, in 2013 and 2015, respectively.\\
Her current research interests include the application of machine learning in social network problems.
\end{IEEEbiography}
\begin{IEEEbiography}[{\includegraphics[width=1in,height=1.25in,clip,keepaspectratio]{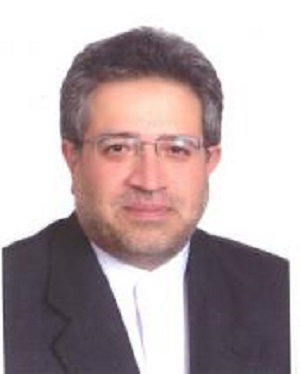}}]{Hamid R. Rabiee}
$(SM’07)$ received his B.S. and M.S. degrees (with great distinction) in electrical engineering from California State University, Long Beach (CSULB), CA, in 1987 and 1989, respectively; the EEE degree in electrical and computer engineering from University of Southern California (USC), Los Angeles, CA; and the Ph.D. degree in electrical and computer engineering from Purdue University, West Lafayette, IN, in 1996. From 1993 to 1996, he was a Member of the Technical Staff with AT$\&$T Bell Laboratories. From 1996 to 1999, he worked as a Senior Software Engineer at Intel Corporation. From 1996 to 2000, he was an Adjunct Professor of electrical and computer engineering at Portland State University, Portland, OR; with Oregon Graduate Institute, Beaverton, OR; and with Oregon State University, Corvallis, OR. Since September 2000, he has been with the Department of Computer Engineering, Sharif University of Technology, Tehran, Iran, where he is currently a Professor and the Director of Sharif University Advanced Information and Communication Technology Research Center (AICT), Digital Media Laboratory (DML), and Mobile Value Added Services Laboratory (VASL). He is also the founder of AICT, Advanced Technologies Incubator (SATI), DML, and VASL. He is a holder of three patents. He has been the Initiator and Director of national and international-level projects in the context of United Nation Open Source Network program and Iran National ICT Development Plan. He has received numerous awards and honors for his industrial, scientific, and academic contributions.
\end{IEEEbiography}
\begin{IEEEbiography}[{\includegraphics[width=1in,height=1.25in,clip,keepaspectratio]{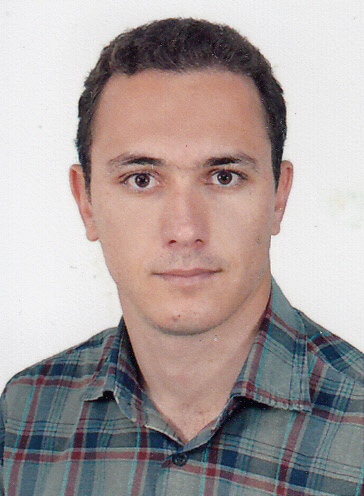}}]{Ali Khodadadi}
received his B.Sc. and the M.Sc. degrees in information technology engineering from Sharif University of Technology, Tehran, Iran, in 2010 and 2012, respectively. He is currently working toward his Ph.D. degree in the Department of Computer Engineering, Sharif University of Technology. His current research interests include using machine learning approaches to solve social and complex networks problems such as inferring networks of diffusion, Bayesian community detection, multilayer network analysis, and user activity modeling over social media sites.
\end{IEEEbiography}








\end{document}